\documentclass{article} 

\PassOptionsToPackage{numbers,sort&compress}{natbib}
\usepackage[preprint]{neurips_2026}

\usepackage[utf8]{inputenc}
\usepackage[T1]{fontenc}
\usepackage{hyperref}
\usepackage{url}
\usepackage{graphicx}
\usepackage{booktabs}
\usepackage{amsfonts}
\usepackage{amsmath}
\usepackage{amssymb}
\usepackage{nicefrac}
\usepackage{microtype}
\usepackage{algorithm}%
\usepackage{algorithmicx}%
\usepackage{algpseudocode}%
\usepackage{wrapfig}
\usepackage{float}

\usepackage{amsmath,amsfonts,bm}

\def\eqref#1{equation~\ref{#1}}

\def\1{\bm{1}}

\DeclareMathAlphabet{\mathsfit}{\encodingdefault}{\sfdefault}{m}{sl}
\SetMathAlphabet{\mathsfit}{bold}{\encodingdefault}{\sfdefault}{bx}{n}

\graphicspath{{../}{../figures/}{figures/}}

\usepackage[x11names]{xcolor}

\definecolor{revision}{RGB}{60,120,255}

\definecolor{ModelBlue}{RGB}{0,0,0}   
\definecolor{HumanGray}{RGB}{0,0,0}  
\definecolor{AnchorTeal}{RGB}{0,0,0}   
\definecolor{PosPurple}{RGB}{0,0,0}   
\definecolor{OddOrange}{RGB}{0,0,0}   
\definecolor{BatchIndigo}{RGB}{0,0,0}  
\definecolor{GradRed}{RGB}{0,0,0}      
\definecolor{MaskBrown}{RGB}{0,0,0}
\definecolor{TripletPink}{RGB}{0,0,0}   
\definecolor{RSAcyan}{RGB}{0,0,0}       

\newcommand{\mcol}[2]{\begingroup\color{#1}#2\endgroup}

\newcommand{\anc}[1]{\mcol{AnchorTeal}{#1}}       
\newcommand{\pos}[1]{\mcol{PosPurple}{#1}}        
\newcommand{\odd}[1]{\mcol{OddOrange}{#1}}        
\newcommand{\Kset}{\mcol{BatchIndigo}{\mathcal{K}}}
\newcommand{\Gset}{\mcol{GradRed}{\mathcal{G}}}
\newcommand{\Mset}{\mcol{MaskBrown}{\mathcal{M}}}
\newcommand{\zv}[1]{\mcol{ModelBlue}{\mathbf{z}_{#1}}}

\newcommand{\Ltrip}{\mcol{TripletPink}{\mathcal{L}_{\text{triplet}}}}
\newcommand{\LRSA}{\mcol{RSAcyan}{\mathcal{L}_{\text{RSA}}}}
\newcommand{\Lhyb}{\mathcal{L}_{\text{hybrid}}}

\title{Behavioral Geometric Supervision Aligns Video Foundation Models with Human Social Perception}

\author{
\begin{tabular*}{0.6\textwidth}{@{\extracolsep{\fill}}lr@{}}
Kathy Garcia$^{1}$ & Leyla Isik$^{1,2}$ \\
\texttt{kgarci18@jhu.edu} & \texttt{lisik@jhu.edu}
\end{tabular*}
\\[2em]
$^1$Department of Cognitive Science \hspace{10pt}
$^2$Department of Biomedical Engineering \\
Johns Hopkins University
}

\begin{document}

\maketitle

\begin{abstract}

Current video foundation models, including the strongest self-supervised models such as V-JEPA2, fail to capture how humans organize social information in dynamic scenes. For example, across a range of diverse vision models tested, none were able to predict human similarity judgments to social video clips as well as a sentence embedding model of the caption text (MPNet). We show this gap in vision model performance can be closed by a compact behavioral supervisory signal. We introduce \emph{behavioral geometric supervision (BGS)}: a hybrid objective that constrains local and global pairwise embedding geometry to match the relational similarity structure across videos. We apply this method using a new human similarity dataset, containing 49,484 odd-one-out judgments from 250 naturalistic social video clips, and low-rank adaptation across four ViT backbones (V-JEPA 2/2.1, TimeSformer, VideoMAE, and CLIP). We find that one of the best fine-tuned models, V-JEPA 2.1, nearly triples in performance compared to the pre-trained baseline and reaches close to the noise ceiling, exceeding the strongest sentence-embedding baseline. In addition, finetuned models (i) capture unique variance in human judgments that caption-based language embeddings do not, (ii) develop interpretable social-affective attributes (valence, arousal, and dominance) despite never being trained on any of these attributes, (iii) zero-shot transfer to a separate dataset of out-of-distribution abstract social interactions, and (iv) shift spatial attention from scene context to socially informative regions (faces, gaze, and interacting bodies). A matched language-distillation control fails to reproduce these gains, ruling out caption transfer as the mechanism. Our results show how a modest amount of human behavioral data can steer video models toward human-like social visual understanding.

\end{abstract}

\section{Introduction}\label{sec:intro}

Humans effortlessly perceive the visual social world with remarkable nuance: we can easily tell whether two people are comforting each other, collaborating, or competing, all by watching brief interactions. As AI systems increasingly interpret and interact in human-centered environments, aligning their representations with human social perception is essential. Yet how this capacity arises in artificial systems, and whether the dynamic relational structure of human social perception can emerge in purely visual models at all, remains unknown.

Recent benchmarks have shown that current vision foundation models systematically under-explain human social judgments compared to language models \citep{garcia2025modeling}, which have been shown to capture multimodal sensory judgments \citep{marjieh_2023}. While prior alignment work has shown that human similarity judgments can reshape vision model representations \citep{muttenthaler_2023_improving, fu_2023_dreamsim, sundaram_when_2024, mahner2025dimensions}, it has not applied these methods to dynamic social videos where similarity depends on temporally integrated information. Recent concurrent work \citep{policzer2025one, schad2025vibe} has "brain-tuned" models on social-cognition using human fMRI data, highlighting the value of cognitive supervision but at infrastructure costs (based on expensive fMRI data) that limit scalability. Two questions therefore remain open: (1) can human social similarity structure be induced in video foundation models through scalable behavioral supervision, and (2) is the resulting structure visual-social or is it redundant with the semantic structure language embeddings already encode \citep{marjieh_2023}?

Here, we introduce \emph{behavioral geometric supervision (BGS)}: a differentiable training signal in which pairwise distances in a video model's embedding space are constrained to match both the local and global relational geometry of human odd-one-out similarity judgments, using a hybrid triplet-RSA objective. We collect a novel dataset of 49,484 similarity judgments from 250 naturalistic two-person social videos, and fine-tune four heterogeneous vision transformer backbones (V-JEPA 2, TimeSformer, VideoMAE, CLIP) with our new hybrid triplet–RSA objective applied through  (LoRA) \citep{hu2022lora}, updating fewer than 2\% of parameters. We then test, by direct comparison against contrastive, predictive, and language-distillation alternatives on the same backbones, whether the resulting structure is reducible to other types of supervision.

\paragraph{Contributions.} We report three main contributions:{\label{par:contributions}}

\textbf{(1) A dynamic social video similarity benchmark reveals missing social-visual representations.}
We release the first large-scale human similarity benchmark for dynamic social video: 49,484 odd-one-out judgments from 245 participants for 250 naturalistic two-person clips. We identify a representation gap in vision models: across 14 pretrained vision encoders (e.g., V-JEPA~2/2.1), none match human social similarity as well as caption-based MPNet embeddings, indicating video pretraining misses social-visual structure.

\textbf{(2) Behavioral Geometric Supervision for compact relational alignment.}
We introduce Behavioral Geometric Supervision (BGS), a LoRA-based hybrid triplet + differentiable RSA objective updating $<$2\% of parameters. BGS closes the gap across four heterogeneous ViT backbones, explaining on average 76.5\% of the split-half ceiling, and 81\% on the strongest model. MPNet distillation, triplet-budget matching, and triplet-only/RSA-only ablations all underperform BGS, showing that the gains require behavioral targets and joint local/global geometric supervision.

\textbf{(3) Transferable, interpretable, capability-preserving social structure.}
BGS-aligned representations transfer zero-shot to out-of-distribution (OOD), abstract, dyadic social interactions, develop interpretable social-affective attributes (valence, arousal, dominance) without supervision, reorganize spatial attention to socially informative regions (faces, gaze, interacting bodies), and preserve action recognition on UCF101. 

\section{Related Work}

We situate our work against three closely related areas: human similarity judgments in vision and alignment of models with human perception, video pretraining beyond category level recognition, and language-based accounts of perceptual similarity. 

\paragraph{Behavioral alignment of Vision models with Human Similarity Judgments.}
Human similarity judgments have long been used to probe mental representations \citep{Biederman1987, Edelman1998, Goldstone1994, Hebart2020, Nosofsky1986} with large-scale odd-one-out (OOO) behavioral studies mapping latent ``similarity spaces'' humans use for objects \citep{Hebart2020}, materials \citep{Schmidt2025}, and interaction environments \citep{Josephs2023}. 
Recent alignment work uses these judgments to reshape model representations: VICE \citep{muttenthaler2022vice}, \citep{muttenthaler_2023_improving}, DreamSim \citep{fu_2023_dreamsim}, and AligNet \citep{Muttenthaler2025} show that human derived similarity signals yield more perceptually aligned and interpretable features. However, most of this work focuses on static images, low-level perceptual comparisons, or synthetic domains, while \citep{dima_simjudg_2022} found that human judgments of dynamic stimuli rely more on social-affective features than surface visual/scene features. Here we collect the first large-scale dataset of social video similarity, and apply new methods in BGS to video models. 

A parallel line of work \citep{kaufman2023RLHF, ovip_2025, adavip_2025} aligns multimodal models to human preferences via scalar rewards, direct preference optimization, or online negative sampling, but optimizes output quality and leaves internal representational geometry largely unconstrained. These methods also  typically require human-in-the-loop annotation at scale \citep{furuta2024RLHF, li2024RLHF}. In contrast, we apply BGS directly shaping the pairwise geometry of video embeddings using relational similarity judgments.

\paragraph{Beyond categorical video pretraining.} 
Modern video encoders, including transformer-based architectures trained at scale (TimeSformer \citep{bertasius_is_2021}, ViViT \citep{arnab2021vivit}, and VideoMAE \citep{tong2022videomae}) have achieved strong results on action classification benchmarks. Yet their dominant objectives and evaluations emphasize category-level recognition (e.g., ``dancing'' vs. ``cooking'') instead of higher-level aspects of social behavior (e.g., intentions, affect, interactions, etc.). Multimodal video-language models such as VideoCLIP \citep{xu2021videoclip}, All-in-One \citep{wang2022allinone}, and Qwen3 \citep{bai2025qwen3} introduce textual supervision, which can support more abstract semantic representations. These approaches still depend on captions and descriptions, which can miss the relational and affective cues people use to compare social scenes.

The V-JEPA family \citep{assran2025vjepa2, murlabadia2026vjepa21} provides a particularly strong test case for our question of vision model representations beyond categorization. V-JEPA 2 \citep{assran2025vjepa2} is a large self-supervised predictive video model trained on roughly 1M hours of video with a joint-embedding predictive objective, reporting state-of-the-art results on action understanding, planning, and robotic manipulation. V-JEPA 2.1 \citep{murlabadia2026vjepa21} extends this line with a dense prediction loss and deep self-supervision, yielding spatially and temporally consistent features that further improve these capabilities. Yet, neither model has been evaluated on whether it can decode higher-level social features from human social similarity structure. We test both V-JEPA 2 ViT-L ($\sim$300M params) and the distilled V-JEPA 2.1 ViT-B variant ($\sim$ 86M params) to compare across different parameter sizes, and ask whether predictive video pretraining alone is sufficient, or behavioral similarity supervision provides distinct alignment.

\paragraph{Language baselines and caption-derived structure.}
\citet{marjieh_2023} demonstrated that caption-based language embeddings approximate human similarity across sensory modalities, often outperforming pretrained vision encoders. As such, we test a range of language model embeddings from video captions as our reference, and treat exceeding the highest performing embedding space, MPNet \citep{song2020mpnet}, as a meaningful bar. Exceeding this baseline suggests behavioral fine-tuning is capturing structure beyond what can be recovered from captions alone. It remains an open question whether and to what extent vision models can surpass this baseline. Here, we address this question using not only performance comparison, but also variance partitioning and a matched language-distillation control absent from earlier behavioral similarity approaches \citep{fu_2023_dreamsim, muttenthaler2022vice, Muttenthaler2025, sundaram_when_2024}.

\section{Methods}\label{methods}


\begin{figure}[hb]
    \vspace{-1.2\baselineskip}
    \centering
    \includegraphics[width=0.75\linewidth]
    {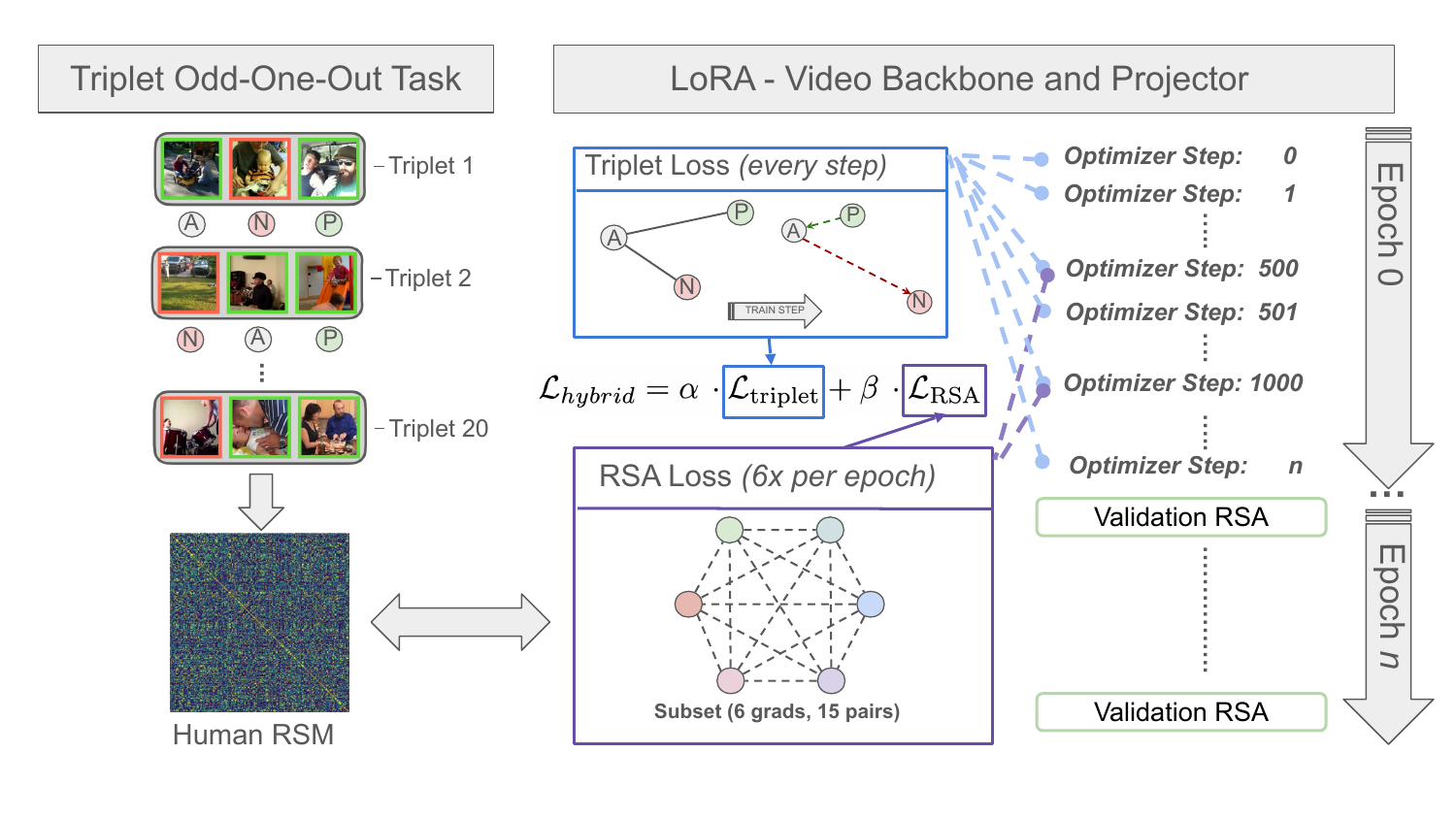}
    \caption{Triplet OOO Dataset \& Video Model Hybrid Fine-Tuning. Human similarity judgments are collected in 20-trial odd-one-out sets and define positive and negative training signals. The model is updated with triplet loss (blue) on Anchor (A), Positive (P), and Negative (N) batches. 6x per epoch, an RSA loss (purple) is applied to a 24-video subset by aligning model pairwise distances with triplet-derived human similarities (see Eq.~\ref{eq:hybridloss} for combined triplet-RSA objective).}
    \label{fig:fig1}
    \vspace{-0.8\baselineskip}
\end{figure}

Our approach has two stages:\\
\textbf{(1)} Measure the human-perceived similarity: we collect odd-one-out judgments on video triplets to construct a human similarity matrix, \\
\textbf{(2)} Leverage behavior-guided fine-tuning on a video model with the objective of matching its embedding distances to human similarity structure. This is achieved through a hybrid loss function that enforces both local triplet constraints and global alignment of the full similarity matrices (Fig. \ref{fig:fig1}).

\clearpage

\subsection{Human Similarity Judgment Dataset}\label{sec:sim_judg_dataset}
We introduce a dense large-scale dataset of human similarity judgments of short social video clips. 

\paragraph{Stimuli.} The stimulus set is 250 3-second clips of everyday two-person social interactions from a curated subset~\citep{mcmahon_hierarchical_2023, garcia2025modeling} of the Moments in Time dataset~\citep{monfortmoments2019}, paired with one-sentence captions used to evaluate language models (Appendix~\S\ref{methods_sentence_captioning}).

\paragraph{Triplet odd-one-out task.} Similarity judgments used a triplet odd-one-out paradigm~\citep{Hebart2020}: on each trial, participants saw three videos and were instructed to ``focus on what the people are doing and choose the odd-one-out,'' implicitly nominating the remaining pair as more similar. To ensure full pairwise coverage with minimal participant load, we generated triplets with a greedy set-cover algorithm (Appendix Alg.~\ref{alg:triplet_selection}) that guarantees every video pair appears in at least one triplet. 245 participants (mean age 19.5; 64\% female; 75\% Asian or White; recruited via a U.S.\ university research platform) completed the study under IRB approval (see Appendix Fig.~\ref{fig:demographics} for full demographics, and \S\ref{sec:limitations} for the discussion on sampling limitations).

\paragraph{Splits and similarity matrices.} We adopt the pre-determined train-test split~\citep{mcmahon_hierarchical_2023}: 200 train videos (24{,}096 triplets) and 50 test videos (368 triplets). Human similarity is estimated as the frequency with which two videos were judged the more-similar pair, yielding a $200\times200$ training RSM and a $50\times50$ test RSM. Crucially, the similarity judgments on the 50 test videos are \emph{stimulus-disjoint} from training (no test stimulus appears in any training triplet) providing a conservative generalization test; the test RSM covers 637 of 1{,}225 possible pairs. We use Spearman correlation squared throughout for robustness to non-uniform sampling density and spearman brown corrected for the split-half reliability noise ceiling of $R^2 = 0.213$ (Appendix~\S\ref{app:splithalf}).

\subsection{Pretrained Vision and Language Models}

We evaluate 12 pretrained video encoders (CNN- and Transformer-based), 2 image encoders (CLIP ViT-B/32, DINO-ViT-B/16), 2 vision-language models (Qwen3, LFM2.5), and 22 sentence-embedding language model baselines spanning diverse architectures and training objectives (full list: Appendix Tab.~\ref{tab:perf-budget}).\footnote{We also include Qwen3-VL-2B-Instruct as a vision-language 
reference point, selected for parameter-count parity with V-JEPA~2 
(Appendix~\S\ref{app:vlm_details}). For each model layer, we extract 
embeddings for all videos, or sentence captions where applicable, compute 
a $50 \times 50$ cosine-similarity matrix, and compare it to the human 
test-set RSM using RSA \citep{kriegeskorte_representational_2008}. 
Embedding dimensions are standardized via Johnson--Lindenstrauss random 
projection with $\varepsilon = 0.1$ (Appendix~\ref{app:supplementary}).} To select the evaluation layer, we perform a 5‑fold cross‑validation on the 200‑video training set across all model layers, choose the layer with the highest mean Spearman’s $\rho$ across folds, and then fix that layer for evaluation on the held‑out 50‑video test set. 

\subsection{Behavior-Guided Fine-Tuning of Vision transformers}\label{sec:beh_guided_fine_tuning}
Our core approach is to fine-tune vision transformers using the human judgments as supervision. To test whether BGS generalizes across pretraining objectives, we apply lightweight \textsc{LoRA} fine-tuning to four backbones spanning four distinct pretraining regimes, updating less than 2\% of parameters while freezing the rest. This approach inserts low-rank matrices into each attention layer (rank = 16). We show that our results generalize across a diverse range of vision transformers, and report fine-tuning results for modern joint-embedding predictive models (V-JEPA-2 ViT-L, \cite{assran2025vjepa2} and V-JEPA-2.1 ViT-B, \cite{murlabadia2026vjepa21}), the highest performing image model (CLIP, \citep{radford_learning_2021}), the highest performing video transformer (TimeSformer \citep{bertasius_is_2021}), and an additional video transformer VideoMAE \citep{tong2022videomae}. 

\subsubsection{Hybrid Loss Function}

Both terms in our loss are instances of BGS: each takes relational judgments as input and produces gradients that adjust pairwise distances in embedding space. We design a loss \(\Lhyb\) that combines a triplet loss term (\(\Ltrip\)) and an RSA loss term (\(\LRSA\)) to address local \& global alignment (Fig.~\ref{fig:fig1}).

\paragraph{Shared notation and distance.}
Let \(f(v)\) be the embedding of video \(v\). We use \(\ell_2\)-normalized embeddings \(\zv{i} = f(v_i)/\|f(v_i)\|_2\) and define a single cosine-distance operator shared by both losses: \\
\begin{equation}
    d(i,j) \;=\; 1 - \langle \zv{i}, \zv{j} \rangle .
    \label{eq:distance}
\end{equation}

\paragraph{Triplet Loss (local constraints)}
For each human odd-one-out judgment we seek to ensure the distance between anchor video \(\anc{i}\) and its positive pair \(\anc{j}\) is less than the distance to its negative pair \(\odd{k}\) (odd-one-out) by a margin of \(\gamma\). Specifically, we penalize violations of a margin \(\gamma=0.2\):

\begin{equation}
    \Ltrip(\anc{i},\pos{j},\odd{k})
    \;=\;
    \max \ \!\bigl\{0,\; d(\anc{i},\pos{j}) - d(\anc{i},\odd{k}) + \gamma \bigr\}.
    \label{eq:triplet}
\end{equation}

\paragraph{RSA Loss (global geometry)}
To shape the broader geometry toward human similarity structure, we inject a step based on representational similarity analysis (RSA, \cite{kriegeskorte_representational_2008}) step six times per epoch. At each RSA step, we sample a batch of \(\Kset{=}24\) videos and designate a subset of \(\Mset{=}6\) indices \(\Gset\subset\Kset\) whose embeddings carry gradients. We limit gradients to \(\Mset{=}6\) to keep memory manageable: each RSA step considers all $\binom{24}{2}=276$ pairwise distances but backpropagates only through the 123 pairs involving at least one gradient carrying video, which empirically provided sufficient supervision without the overhead of updating all \(24\) embeddings.

We calculate model RSM entries with \(d(\cdot,\cdot)\) for all unordered pairs \(\{i,j\}\subset\Kset\) with \(i\neq j\) and \(i\in\Gset\) or \(j\in\Gset\). Corresponding human distances \(d^{\text{H}}(i,j)\) are taken from the split-specific behavior RSM, masking out pairs without judgments to create a masked index set \(\Mset\). 

The RSA loss is the negative RSA score between the \(z\)-scored model and human distances of the masked index set \(\Mset\):
\begin{equation}
    \LRSA
    \;=\;
    -\,\mathrm{corr} \ \!\Big(
        z \ \!\big(\operatorname{vec}(d)\big)[\Mset],\;
        z \ \!\big(\operatorname{vec}(d^{\mathrm H})\big)[\Mset]
    \Big),
    \label{eq:rsa}
\end{equation}

where \(\operatorname{vec}(\cdot)\) denotes vectorization of the upper triangle, and \(z(\cdot)\) denotes per-step standardization to zero mean and unit variance. 

Pearson correlation is used for the RSA loss during training to ensure the loss is differentiable. 

\paragraph{Hybrid Loss.}
We combine triplet (local) and RSA (global) supervisions with a weighted objective:
\begin{equation}
\Lhyb^{(t)}
=
\alpha\,\Ltrip^{(t)}
+
I_{\mathrm{RSA}}(t)\,
\beta^{(t)}\,\LRSA^{(t)} ,
\label{eq:hybridloss}
\end{equation}

where $\Ltrip$ captures fine-grained constraints from odd-one-out judgments and $\LRSA$ encourages broader geometric alignment on sampled subsets. The indicator $I_{\mathrm{RSA}}(t)$ equals 1 when RSA supervision is applied at optimizer step $t$, and 0 otherwise. Specifically, we apply the RSA loss at six evenly spaced optimizer steps per epoch. We fix $\alpha = 0.7$ and linearly ramp $\beta^{(t)}$ from 0.3 to 0.7 over training epochs. We do this to emphasize the local triplet loss early and ensure the model starts by getting the odd-one-out relationships correct, then gradually increase the weight of the global RSA loss, $\beta^{(t)}$, as training progresses to fine-tune the overall similarity structure (see Appendix \S\ref{app:hybrid_loss}). 

\paragraph{Training Procedure.}
We fine-tune for $50$ epochs with AdamW \citep[see][]{AdamW_2017} with learning rate = $1\times10^{-4}$, mixed precision, and gradient-checkpointing, using a batch size of $4$. At each optimizer step, we apply the triplet loss; the RSA term is injected periodically as described above. We select the best checkpoint based on RSA validation performance on a held-out $20\%$ split of the training judgments (monitoring explained variance $R^2$). For ablations, we also train models with triplet-only and RSA-only objectives under the same optimizer and schedule.

\paragraph{Language-distillation control}
We also train a matched language-distillation control that replaces human triplet judgments with similarity of caption embeddings from MPNet (the highest-performing language model) as the supervisory target. Per-video captions are concatenated and encoded (L2-normalized outputs $z_i$), yielding $D^{\mathrm{MPNet}}_{ij} = 1 - z_i^\top z_j$ over the 200 training videos. We optimize the same hybrid loss as the behavioral run (Eqs. \ref{eq:triplet} -- \ref{eq:rsa}) with same weights as hybrid run, substituting $D^{\mathrm{MPNet}}$ for $d^{\mathrm{H}}(i,j)$ in the RSA term and synthesizing odd-one-out triplets directly from MPNet space, mirroring the human odd-one-out task with MPNet as the rater. All other hyperparameters match the behavioral hybrid run, isolating the supervision target as the only difference.

\subsubsection{Linear-Probe Evaluations: Social-Affective Generalization \& Action Recognition}\label{subsec:auxiliary_eval}
To assess whether human similarity alignment changes model representations beyond the primary similarity task, we conducted two auxiliary linear-probe evaluations. First, we tested generalization to five social-affective attributes labeled in the original video dataset \citep{mcmahon_hierarchical_2023}: {\em Intimacy} (how intimate or personal the interaction is), {\em Valence} (overall emotional positivity vs negativity), {\em Arousal} (energy or intensity of the action), {\em Dominance} (power dynamics between people), and {\em Communicating} (whether people in the video are communicating with one another). They were independently rated by and averaged across annotators, then $z$-scored across the 250 videos. We trained ridge-regression linear probes on layer-wise model embeddings using the same train-test split as in the main experiments. Second, to test whether human-aligned fine-tuning preserved the models' original action-recognition capacity, we evaluated frozen pretrained and fine-tuned video backbones on UCF101 \citep{soomro2012ucf101}. We extracted model embeddings and trained linear probes on UCF101 split1 across three random seeds, reporting Top-1 accuracy as mean$\pm$sd (see Appendix~\ref{app:ucf101}).

\section{Results}\label{experiments_results}

\subsection{Pretrained vision models are worse at matching human video similarity than language model embeddings of video captions. }

\begin{wrapfigure}[26]{r}{0.65\textwidth}
    \centering
    \vspace*{-0.4cm}
    \includegraphics[width=\linewidth]{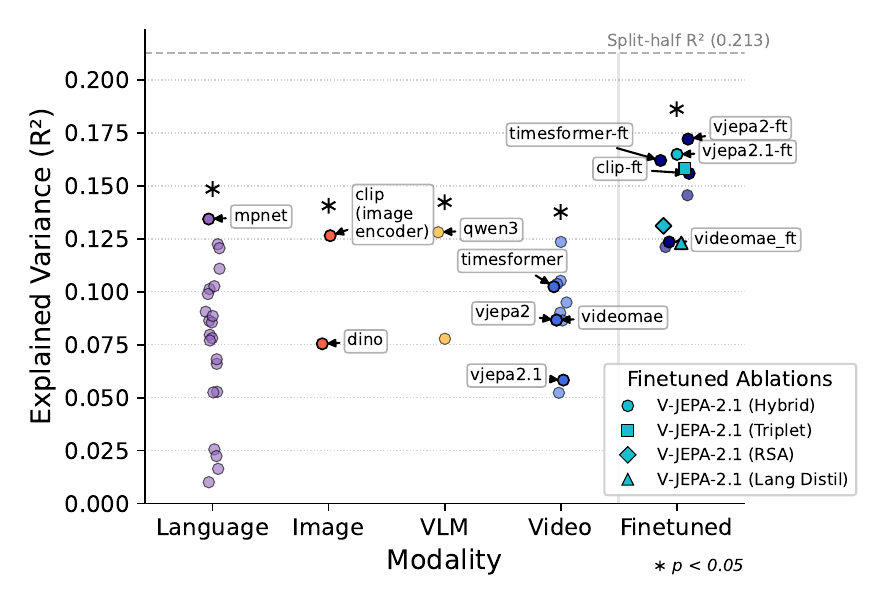}
    \vspace{-0.48cm}
    \caption{Explained variance between model and human representational geometries. Layer-wise model RSMs were compared with the human similarity RSM using Spearman RSA, reported as $R^2$. Shapes denote V-JEPA-2.1 fine-tuning ablations; the dashed line shows the human split-half noise ceiling ($R^2 = 0.213$; see \S\ref{app:splithalf}). Asterisks indicate significance within modality with p < 0.05 across 1000 permutations.}
\label{fig:fig2}
\end{wrapfigure}

Across twelve pretrained video encoders, two image encoders, two vision-language models, and 22 caption-based language encoders, pretrained vision models all provided a worse match to human social similarity than the best sentence encoder (\textit{paraphrase-multilingual-mpnet-base-v2}, $  R^2 = 0.134  $; OOO accuracy 70.38\%; Figure \ref{fig:fig2}). Due to the high performance, we consider these sentence embeddings as a baseline going forward.
Among pretrained baselines, the best video model (X3D-M) reaches only $  R^2 = 0.124  $; OOO accuracy 68.48\%;  V-JEPA-2.1 ViT-B (distilled) or ViT-L (large) only achieve $R^2 = 0.058$; OOO accuracy 57.60\%; and $R^2 = 0.086$; OOO accuracy 63.85\% respectively (Appendix~Tab.~\ref{tab:perf-budget}).
This pattern accords with prior work \citep{marjieh_2023} showing that language embeddings approximate human similarity across perceptual domains. Yet participants performed a purely visual task without captions, exposing critical gaps in current pretrained video representations.

\subsection{Behavioral fine-tuning induces human-like social similarity structure in video models}\label{sec:beh_fine_tuning}

We next ask whether we can imbue video models with more human-like similarity structure via fine-tuning. After fine-tuning, the strongest fine-tuned model, V-JEPA 2 ViT-L reached $ R^2 = 0.172$, exceeding MPNet ($ R^2 = 0.134 $). The distilled V-JEPA 2.1 ViT-B variant, also exceeds the MPNet baseline, reaching $  R^2 = 0.165  $, equivalent to 78\% of the split-half noise ceiling ($  R^2 = 0.213  $; Fig.~\ref{fig:fig2}). Fine-tuning with the hybrid loss function outperforms both the local triplet ($  R^2 = 0.158  $) and global RSA loss ($  R^2 = 0.131  $) alone, and all BGS finetuned models outperform the language distillation control ($  R^2 = 0.123  $). The same hybrid procedure produces consistent gains across all four backbones (TimeSformer, VideoMAE, CLIP, V-JEPA 2/2.1), spanning supervised classification, masked reconstruction, contrastive vision-language, and self-supervised predictive pretraining, while fully preserving action recognition performance with no catastrophic forgetting (Appendix~\ref{app:ucf101}). BGS hence recovers human-aligned social structure regardless of pretraining objective. 

The test set RSM contains judgments for 637/1{,}225 pairs. Bootstrap resampling of the 637 
observed pairs (10{,}000 resamples) confirms the fine-tuning gain is robust: $R^2$ rises from $0.043$ $[95\%\;\mathrm{CI}:\; 0.016,\;0.080]$ (pretrained V-JEPA 2.1 ViT-B) to $0.160$ $[0.109,\;0.216]$ (hybrid fine-tuned), with paired 
$\Delta\ R^2 = +0.118$ $[+0.065,\;+0.173]$, $p < 0.0001$ (see Appendix Tab.~\ref{tab:perf-budget}).

\subsection{Finetuning captures social structure seen in language models and beyond.}\label{finetuning_social_structure}

\begin{wrapfigure}[24]{r}{0.48\textwidth}
    \centering
    \includegraphics[width=0.46\textwidth]{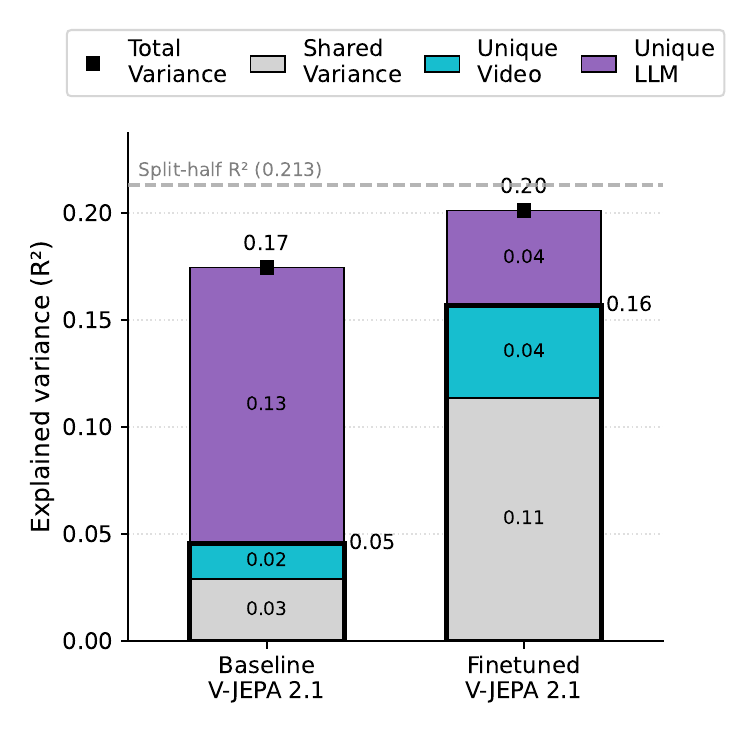}
    \vspace{-1.2\baselineskip}
    \vspace*{0.11cm}
    \caption{
Variance partitioning between V-JEPA 2.1 and MPNet before and after finetuning. 
Finetuning shifts explained variance from language-specific to shared and 
video-specific structure, bringing total variance near the split-half reliability 
ceiling. Black outlines indicate total variance explained by video-model.
    }
    \label{fig:fig3}
\end{wrapfigure} 

Variance partitioning using the best language embedding as a reference further supports the strength of our fine-tuning approach (Fig.~\ref{fig:fig3}). We fit multiple regressions predicting human similarity distances from model distances and decomposed the $R^2$ contributions of the pretrained V-JEPA 2.1, the fine-tuned model, and the language embedding. In the pretrained baseline, the video model contributed little unique variance ($R^2=0.02$) beyond what overlapped with the language model ($R^2=0.03$). After hybrid fine-tuning, the partitioning shifts in two distinct ways. First, structure that was uniquely accessible to language pre-tuning ($R^2=0.13$) collapses into the shared component ($R^2=0.11$), indicating the fine-tuned video model now expresses most of what caption embeddings uniquely encoded. Second, video-unique variance doubles from $R^2=0.02$ to $0.04$; small in absolute terms, but persistent: the fine-tuned model captures social-similarity structure that even an idealized caption embedding cannot reach. Total explained variance approaches the ceiling, leaving little headroom for further alignment. We also observe this trend with other fine-tuned video models (Appendix Fig.~\ref{fig:model_timesformer_partitioning}).

\subsection{OOD generalization to abstract social interactions with no catastrophic forgetting}\label{sec:recovered_structure}

A central question is whether BGS learns a representation that is merely better at the training benchmark or recovers broader social structure. We test this in three complementary ways: social-affective linear probes, cross-domain transfer to abstract social interactions, and retention of standard action-recognition performance.

\paragraph{Interpretable social-affective attributes recovered.}
We tested whether BGS reorganizes representations along human-interpretable social-affective dimensions despite never receiving attribute labels. Following \citep{mcmahon_hierarchical_2023}, we trained ridge-regression probes on five densely annotated attributes: communication, valence, arousal, intimacy, and dominance. BGS substantially improved over pretrained baselines on arousal, dominance, and valence, with modest gains on intimacy and a small decline on communication (Appendix~\ref{app:beh_encoding}). Because they never appear in the supervision signal, such gains indicate that human similarity judgments encode an interpretable social-affective geometry that BGS makes linearly accessible. A matched language-distillation control failed to reproduce these gains on 4/5 attributes; the sole exception is communication, where MPNet's advantage is unsurprising given that words like ``talking'' and ``discussing'' are often directly in caption text.

\paragraph{Zero-shot transfer to abstract social interactions.}
We next tested whether BGS-induced structure transfers beyond the naturalistic training distribution. We evaluated the frozen BGS-aligned V-JEPA 2.1 ViT-B encoder on PHASE~\citep{netanyahu_phase_2021}, a dataset of two-agent interactions, rendered as geometric shapes (no faces, bodies, and minimal scene context) moving in a 2D physical world (Appendix Fig.~\ref{fig:phase_stimuli}), which presents a strong case of OOD generalization relative to the natural videos in model pre-training or fine-tuning. Following~\citet{malik_relational_2023}, a logistic-regression probe trained on 400 videos and evaluated on the 100-video held-out generalization set reaches \textbf{81\% accuracy} (Wilson 95\% CI: 0.72--0.87), a 26-point gain over the pretrained baseline (55\%), matching average human agreement ($\sim$0.83) within sampling error (Appendix Fig.~\ref{fig:phase_summary}). To our knowledge, this is the first demonstration that pixel-level video representations can reach prior baselines on this benchmark.

\paragraph{Action recognition is preserved.}
Finally, we tested whether BGS sacrifices generic action-recognition capacity. We evaluated frozen BGS-aligned and pretrained backbones on UCF101 split1 \citep{soomro2012ucf101} using linear probes across three seeds. V-JEPA 2.1 improved slightly after BGS alignment (78.17 $\pm$ 0.03\% $\rightarrow$ 80.31 $\pm$ 0.06\% Top-1; $\Delta=+2.15$ pp), while TimeSformer was essentially unchanged (95.75 $\pm$ 0.18\% $\rightarrow$ 95.70 $\pm$ 0.14\%; $\Delta=-0.05$ pp). BGS therefore preserves, and for V-JEPA 2.1 marginally improves, generic action-recognition performance. Together, these results show that BGS does not merely improve fit to the trained similarity task. It produces representations that generalize to interpretable social-affective judgments and to abstract OOD social interactions, while preserving the discriminative capacities acquired during pretraining. 

\subsection{Spatial attention reorganization toward socially informative regions}\label{sec:attention}
To examine how fine-tuning changes video model representations, we performed attention rollout on V-JEPA~2.1 ViT-B (Appendix Fig.~\ref{fig:attention-comparison}); the pretrained model concentrates spatial attention on a single focal region per frame (e.g., one agent's torso), while fine-tuning redistributes attention across both agents, gaze directions, interacting hands, and shared activity regions. Across all 250 videos, a mean attention divergence of $1 - \bar{r} = 0.18 \pm 0.03$ (mean $\pm$ sd), indicating systematic spatial reorganization rather than isolated shifts. Qualitative triplet examples further show that fine-tuning agrees on odd-one-out judgments with the human choice while the pretrained model disagreed, prioritizing human-relevant cues (Appendix Figs.~\ref{fig:model_human_comparison-single}--\ref{fig:model_human_comparison}). Together with the variance-partitioning result (\S\ref{finetuning_social_structure}) and the cross-domain transfer (\S\ref{sec:recovered_structure}), this spatial reorganization provides a third, mechanistically grounded line of evidence that BGS recovers genuinely social-visual structure rather than statistical regularities of the training corpus.

\section{Discussion}\label{sec:discussion}
Here we show that human social perception, measured via similarity judgments on dynamic social clips, can be induced in video representations through lightweight behavioral fine-tuning, BGS. Our fine-tuned model develops interpretable social features (e.g., valence, arousal, intimacy, dominance) without any direct supervision, captures variance in human judgments that caption-based language embeddings cannot, and reorganizes its spatial attention toward socially informative regions. Our novel objective generalizes across four backbones, and preserves standard action recognition performance.

Our results reposition social perception in video models as a latent property rather than a learned task. Pretrained video backbones, optimized for action classification or masked reconstruction, on their own fail to match human social similarity judgments, but carry enough information about the social world that a compact behavioral supervision signal suffices to make that information usable. 

\subsection{Behavioral relational similarity as latent-structure supervision}
BGS occupies a niche in the space of human aligned training signal: unlike preference learning, it shapes representations rather than outputs and does not require a human in-the-loop; unlike attribute labeling, it captures the integrated structure of human judgment without enumerating individual axes; and unlike contrastive learning, which gets positive and negative pairs from augmentations/category labels, it draws its structure directly from human behavior. This approach is complementary to recent brain-tuning approaches for social video \citep{policzer2025one, schad2025vibe}, while triplet judgments provide a scalable behavioral target for the relational structure people can explicitly compare.

This framework extends naturally beyond social perception: any domain with human similarity data (e.g., objects, faces, scenes, materials, sounds) admits the same supervision, and large-scale triplet datasets already exist in several of these domains \citep{muttenthaler_2023_improving, Josephs2023, fu_2023_dreamsim, Schmidt2025}. 

\subsection{What our behavioral fine-tuning accesses that language models do not}
Fine-tuned video models explain unique human similarity judgments variance beyond the strongest caption based language encoder, and simultaneously increase the variance shared with language (Fig. \ref{fig:fig3}). More unique variance in video embeddings and increased shared variance combined show the signature of a model that has acquired visual social structure that humans rely on \citep{mcmahon_seeing_2023} rather than distilled language: pure distillation would increase shared variance while leaving unique video variance flat/shrinking it. One question naturally arises: why does the language baseline outperform all video models in the first place? Understanding social interactions often requires abstract inferences (goals, roles, affect) that go beyond form and motion. Video models, trained mainly for action classification, may emphasize kinematics and object cues, while caption-based language embeddings encode high-level semantics (e.g., “friends boxing for fun” vs. “strangers fighting angrily”). This makes the increased unique-variance for fine-tuned video models particularly compelling, suggesting humans encode aspects of social structure visually beyond verbal descriptions. 

Three convergent lines of evidence support this: (1) On triplets where the pretrained video model disagrees with humans but the language model agrees, the fine-tuned video model flips its judgments toward the human choice (e.g., prioritizing social relational over surface visual cues; Appendix \S\ref{app:supplemental_comparison}). (2) Spatial attention reorganizes from scene context/background toward faces, gaze, and interactions after fine-tuning (Appendix Fig.~\ref{fig:attention-comparison}), without any spatial supervision during training. (3) Linear probes recover interpretable axes of valence, arousal, and dominance from the fine-tuned representations despite none of them appearing in the supervision signal (Appendix Fig.~\ref{behavior_encoding}). Together, they show that the fine-tuned model has acquired high-level social content, not just richer visual statistics.

\subsection{Limitations}\label{sec:limitations}

Our dataset, though diverse in action categories, draws from a single source corpus (Moments in Time, \citet{monfortmoments2019}) and reflects predominantly Western social norms: all 245 participants were recruited through a U.S.\ university platform, and captions were provided by native English speakers from Prolific. Because social similarity judgments vary with cultural background \citep{Pang2024}, the learned structure should not be assumed to generalize across all populations without cross-cultural validation. Our evaluation focuses on similarity alignment; we test generalization along multiple axes including cross-domain transfer (\S\ref{sec:recovered_structure}) and cross-task transfer (Appendix~\S\ref{app:ucf101}). The aggregate-consensus nature of our similarity data also smooths over individual differences; future work on personalized or subgroup-specific alignment would test the limits of the current approach. Finally, the 250-video stimulus set, while behaviorally densely sampled, is small relative to image-domain benchmarks; cross-dataset replication is the natural next step.

\subsection{Broader Impact}
Video models that are aligned with human social similarity judgments provide a path to more trustworthy and intuitive AI systems. Embeddings that align with human behavior may improve interpretability, video retrieval, and recommendation by way of organizing content that is reflective of human categorization. Our findings suggest that such alignment also promotes representations of social-affective features, with potential applications in affective computing and safety-sensitive domains. However, models that reflect human perception may also inherit human biases. While our dataset is diverse, culturally specific notions of similarity may also be encoded as a result of the aforementioned factors. This motivates future broader studies that include analysis for bias with more diverse annotation sources, ensuring fairness and robustness across populations. 

\section{Conclusion}\label{conclusion}

 Together, these results reposition social perception as a latent property of video representations that becomes accessible only with the right supervision signal, and suggest that relational behavioral data offer a form of alignment different from the preference learning and cross-modal distillation that currently dominate multimodal training. We release our benchmark, trained adapters, and training code as  resources for the community.

\begin{ack}
This work was funded in part by NSF GRFP DGE-2139757 awarded to K.G. and NIMH R01MH132826 awarded to L.I.
\end{ack}

\bibliographystyle{plainnat}
\bibliography{neurips_2026}

@misc{adavip_2025,
  doi = {10.48550/ARXIV.2504.15619},
  url = {https://arxiv.org/abs/2504.15619},
  author = {Lu,  Jinda and Li,  Jinghan and Gao,  Yuan and Wu,  Junkang and Wu,  Jiancan and Wang,  Xiang and He,  Xiangnan},
  title = {AdaViP: Aligning Multi-modal LLMs via Adaptive Vision-enhanced Preference Optimization},
  publisher = {arXiv},
  year = {2025},
  copyright = {arXiv.org perpetual,  non-exclusive license}
}

@misc{marjieh_2023,
  doi = {10.48550/ARXIV.2302.01308},
  url = {https://arxiv.org/abs/2302.01308},
  author = {Marjieh,  Raja and Sucholutsky,  Ilia and van Rijn,  Pol and Jacoby,  Nori and Griffiths,  Thomas L.},
  title = {Large language models predict human sensory judgments across six modalities},
  publisher = {arXiv},
  year = {2023},
  copyright = {Creative Commons Attribution 4.0 International}
}

@misc{ovip_2025,
  doi = {10.48550/ARXIV.2505.15963},
  url = {https://arxiv.org/abs/2505.15963},
  author = {Liu,  Shujun and Wang,  Siyuan and Li,  Zejun and Wang,  Jianxiang and Zeng,  Cheng and Wei,  Zhongyu},
  title = {OViP: Online Vision-Language Preference Learning for VLM Hallucination},
  publisher = {arXiv},
  year = {2025},
  copyright = {Creative Commons Attribution 4.0 International}
}

@misc{AdamW_2017,
  doi = {10.48550/arxiv.1711.05101},
  url = {https://arxiv.org/abs/1711.05101},
  author = {Loshchilov,  Ilya and Hutter,  Frank},
  title = {Decoupled Weight Decay Regularization},
  publisher = {arXiv},
  year = {2017},
  copyright = {arXiv.org perpetual,  non-exclusive license}
}

@inproceedings{hu2022lora,
  title     = {LoRA: Low-Rank Adaptation of Large Language Models},
  author    = {Hu, Edward J. and Shen, Yelong and Wallis, Phillip and Allen-Zhu, Zeyuan and Li, Yuanzhi and Wang, Shean and Wang, Lu and Chen, Weizhu},
  booktitle = {International Conference on Learning Representations (ICLR)},
  year      = {2022},
  url       = {https://arxiv.org/abs/2106.09685},
  doi       = {10.48550/arXiv.2106.09685}
}

@article{mahner2025dimensions,
  title   = {Dimensions underlying the representational alignment of deep neural networks with humans},
  author  = {Mahner, Florian P. and Muttenthaler, Lukas and G{\"u}{\c{c}}l{\"u}, Umut and Hebart, Martin N.},
  journal = {Nature Machine Intelligence},
  year    = {2025},
  volume  = {7},
  number  = {6},
  pages   = {848--859},
  doi     = {10.1038/s42256-025-01041-7}
}

@article{Josephs2023,
  title = {Dimensions underlying human understanding of the reachable world},
  volume = {234},
  ISSN = {0010-0277},
  url = {http://dx.doi.org/10.1016/j.cognition.2023.105368},
  DOI = {10.1016/j.cognition.2023.105368},
  journal = {Cognition},
  publisher = {Elsevier BV},
  author = {Josephs,  Emilie L. and Hebart,  Martin N. and Konkle,  Talia},
  year = {2023},
  month = may,
  pages = {105368}
}

@article{Schmidt2025,
  title = {Core dimensions of human material perception},
  volume = {122},
  ISSN = {1091-6490},
  url = {http://dx.doi.org/10.1073/pnas.2417202122},
  DOI = {10.1073/pnas.2417202122},
  number = {10},
  journal = {Proceedings of the National Academy of Sciences},
  publisher = {Proceedings of the National Academy of Sciences},
  author = {Schmidt,  Filipp and Hebart,  Martin N. and Schmid,  Alexandra C. and Fleming,  Roland W.},
  year = {2025},
  month = mar 
}

@misc{li2024RLHF,
  doi = {10.48550/ARXIV.2410.05677},
  url = {https://arxiv.org/abs/2410.05677},
  author = {Li,  Jiachen and Long,  Qian and Zheng,  Jian and Gao,  Xiaofeng and Piramuthu,  Robinson and Chen,  Wenhu and Wang,  William Yang},
  title = {T2V-Turbo-v2: Enhancing Video Generation Model Post-Training through Data,  Reward,  and Conditional Guidance Design},
  publisher = {arXiv},
  year = {2024},
  copyright = {Creative Commons Attribution 4.0 International}
}

@misc{furuta2024RLHF,
  doi = {10.48550/ARXIV.2412.02617},
  url = {https://arxiv.org/abs/2412.02617},
  author = {Furuta,  Hiroki and Zen,  Heiga and Schuurmans,  Dale and Faust,  Aleksandra and Matsuo,  Yutaka and Liang,  Percy and Yang,  Sherry},
  title = {Improving Dynamic Object Interactions in Text-to-Video Generation with AI Feedback},
  publisher = {arXiv},
  year = {2024},
  copyright = {Creative Commons Attribution 4.0 International}
}

@misc{kaufman2023RLHF,
  doi = {10.48550/ARXIV.2312.14925},
  url = {https://arxiv.org/abs/2312.14925},
  author = {Kaufmann,  Timo and Weng,  Paul and Bengs,  Viktor and H\"{u}llermeier,  Eyke},
  title = {A Survey of Reinforcement Learning from Human Feedback},
  publisher = {arXiv},
  year = {2023},
  copyright = {Creative Commons Attribution 4.0 International}
}

@article{assran2025vjepa2,
  title     = {V-JEPA 2: Self-Supervised Video Models Enable Understanding, Prediction and Planning},
  author    = {Assran, Mahmoud and Alayrac, Jean-Baptiste and Caron, Mathilde and Misra, Ishan and Mialon, Gr{\'e}goire and Bojanowski, Piotr and Joulin, Armand and Synnaeve, Gabriel and Lenc, Karel and Owen, David and Laptev, Ivan and Schmid, Cordelia and Vedaldi, Andrea and Zisserman, Andrew and LeCun, Yann and Touvron, Hugo and Jegou, Herv{\'e}},
  journal   = {arXiv preprint arXiv:2506.09985},
  year      = {2025},
  url       = {https://arxiv.org/abs/2506.09985}
}

@inproceedings{garcia2025modeling,
  title={Modeling dynamic social vision reveals gaps between deep learning and the humans},
  author={Garcia, Kathy and McMahon, Emalie and Conwell, Colin and Bonner, Michael F. and Isik, Leyla.},
  booktitle={Proceedings of the Thirteenth International Conference on Learning Representations},
  year={2025},
  url={https://proceedings.iclr.cc/paper_files/paper/2025/file/b1bdb0f22c9748203c62f29aa297ac57-Paper-Conference.pdf}
}

@misc{soomro2012ucf101,
  doi = {10.48550/arxiv.1212.0402},
  url = {https://arxiv.org/abs/1212.0402},
  author = {Soomro,  Khurram and Zamir,  Amir Roshan and Shah,  Mubarak},
  title = {UCF101: A Dataset of 101 Human Actions Classes From Videos in The Wild},
  publisher = {arXiv},
  year = {2012},
  copyright = {arXiv.org perpetual,  non-exclusive license}
}

@inproceedings{arnab2021vivit,
  title     = {ViViT: A Video Vision Transformer},
  author    = {Arnab, Anurag and Dehghani, Mostafa and Heigold, Georg and Sun, Chen and Lu{\v{c}}i{\'c}, Mario and Schmid, Cordelia},
  booktitle = {Proceedings of the IEEE/CVF International Conference on Computer Vision (ICCV)},
  year      = {2021},
  pages     = {6836--6846},
  publisher = {IEEE},
  address   = {Montreal, Canada},
  doi       = {10.1109/ICCV48922.2021.00676}
}

@inproceedings{tong2022videomae,
  title     = {VideoMAE: Masked Autoencoders are Data-Efficient Learners for Self-Supervised Video Pre-Training},
  author    = {Zhan Tong and Yibing Song and Jue Wang and Limin Wang},
  booktitle = {Advances in Neural Information Processing Systems (NeurIPS)},
  volume    = {35},
  year      = {2022},
  pages     = {3487--3501},
  doi       = {10.5555/3600270.3601002},
  eprint    = {arXiv:2203.12602},
  archivePrefix = {arXiv},
  primaryClass = {cs.CV}
}

@inproceedings{xu2021videoclip,
  title     = {VideoCLIP: Contrastive Pre-training for Zero-shot Video-Text Understanding},
  author    = {Hu Xu and Gargi Ghosh and Po-Yao Huang and Dmytro Okhonko and Armen Aghajanyan and Florian Metze and Luke Zettlemoyer and Christoph Feichtenhofer},
  booktitle = {Proceedings of the 2021 Conference on Empirical Methods in Natural Language Processing (EMNLP)},
  year      = {2021},
  pages     = {6787--6800},
  doi       = {10.18653/v1/2021.emnlp-main.544},
  url       = {https://aclanthology.org/2021.emnlp-main.544/}
}

@article{murlabadia2026vjepa21,
  title   = {V-JEPA 2.1: Unlocking Dense Features in Video Self-Supervised Learning},
  author  = {Mur-Labadia, Lorenzo and Muckley, Matthew and Bar, Amir and Assran, Mahmoud and Sinha, Koustuv and Rabbat, Michael and LeCun, Yann and Ballas, Nicolas and Bardes, Adrien},
  journal = {arXiv preprint arXiv:2603.14482},
  year    = {2026}
}

@article{bai2025qwen3,
  title   = {Qwen3-VL Technical Report},
  author  = {Bai, Shuai and Cai, Yuxuan and Chen, Ruizhe and Chen, Keqin and Chen, Xionghui and Cheng, Zesen and Deng, Lianghao and Ding, Wei and Gao, Chang and Ge, Chunjiang and Ge, Wenbin and Guo, Zhifang and Huang, Qidong and Huang, Jie and Huang, Fei and Hui, Binyuan and Jiang, Shutong and Li, Zhaohai and Li, Mingsheng and Li, Mei and Li, Kaixin and Lin, Zicheng and Lin, Junyang and Liu, Xuejing and Liu, Jiawei and Liu, Chenglong and Liu, Yang and Liu, Dayiheng and Liu, Shixuan and Lu, Dunjie and Luo, Ruilin and Lv, Chenxu and Men, Rui and Meng, Lingchen and Ren, Xuancheng and Ren, Xingzhang and Song, Sibo and Sun, Yuchong and Tang, Jun and Tu, Jianhong and Wan, Jianqiang and Wang, Peng and Wang, Pengfei and Wang, Qiuyue and Wang, Yuxuan and Xie, Tianbao and Xu, Yiheng and Xu, Haiyang and Xu, Jin and Yang, Zhibo and Yang, Mingkun and Yang, Jianxin and Yang, An and Yu, Bowen and others},
  journal = {arXiv preprint arXiv:2511.21631},
  year    = {2025}
}

@article{wang2022allinone,
  title     = {All in One: Exploring Unified Video-Language Pre-training},
  author    = {Wang, Alex Jinpeng and Ge, Yixiao and Yan, Rui and Ge, Yuying and Lin, Xudong and Cai, Guanyu and Wu, Jianping and Shan, Ying and Qie, Xiaohu and Shou, Mike Zheng},
  journal   = {arXiv preprint arXiv:2203.07303},
  year      = {2022},
  doi       = {10.48550/arXiv.2203.07303}
}

@article{Pang2024,
  title = {Cross-cultural Differences in Using Nonverbal Behaviors to Identify Indirect Replies},
  volume = {48},
  ISSN = {1573-3653},
  url = {http://dx.doi.org/10.1007/s10919-024-00454-z},
  DOI = {10.1007/s10919-024-00454-z},
  number = {2},
  journal = {Journal of Nonverbal Behavior},
  publisher = {Springer Science and Business Media LLC},
  author = {Pang,  Hio Tong and Zhou,  Xiaolin and Chu,  Mingyuan},
  year = {2024},
  month = feb,
  pages = {323–344}
}

@article{Conwell2024_deepjuice,
  title = {A large-scale examination of inductive biases shaping high-level visual representation in brains and machines},
  volume = {15},
  ISSN = {2041-1723},
  url = {http://dx.doi.org/10.1038/s41467-024-53147-y},
  DOI = {10.1038/s41467-024-53147-y},
  number = {1},
  journal = {Nature Communications},
  publisher = {Springer Science and Business Media LLC},
  author = {Conwell,  Colin and Prince,  Jacob S. and Kay,  Kendrick N. and Alvarez,  George A. and Konkle,  Talia},
  year = {2024},
  month = oct 
}

@misc{feichtenhofer_2020_x3d,
  doi = {10.48550/ARXIV.2004.04730},
  url = {https://arxiv.org/abs/2004.04730},
  author = {Feichtenhofer,  Christoph},
  title = {X3D: Expanding Architectures for Efficient Video Recognition},
  publisher = {arXiv},
  year = {2020},
  copyright = {arXiv.org perpetual,  non-exclusive license}
}

@article{Biederman1987,
  title = {Recognition-by-components: A theory of human image understanding.},
  volume = {94},
  ISSN = {0033-295X},
  url = {http://dx.doi.org/10.1037/0033-295x.94.2.115},
  DOI = {10.1037/0033-295x.94.2.115},
  number = {2},
  journal = {Psychological Review},
  publisher = {American Psychological Association (APA)},
  author = {Biederman,  Irving},
  year = {1987},
  month = apr,
  pages = {115–147}
}

@article{Edelman1998,
  title = {Representation is representation of similarities},
  volume = {21},
  ISSN = {1469-1825},
  url = {http://dx.doi.org/10.1017/s0140525x98001253},
  DOI = {10.1017/s0140525x98001253},
  number = {4},
  journal = {Behavioral and Brain Sciences},
  publisher = {Cambridge University Press (CUP)},
  author = {Edelman,  Shimon},
  year = {1998},
  month = aug,
  pages = {449–467}
}

@article{Nosofsky1986,
  title = {Attention,  similarity,  and the identification–categorization relationship.},
  volume = {115},
  ISSN = {0096-3445},
  url = {http://dx.doi.org/10.1037/0096-3445.115.1.39},
  DOI = {10.1037/0096-3445.115.1.39},
  number = {1},
  journal = {Journal of Experimental Psychology: General},
  publisher = {American Psychological Association (APA)},
  author = {Nosofsky,  Robert M.},
  year = {1986},
  pages = {39–57}
}

@article{Goldstone1994,
  title = {The role of similarity in categorization: providing a groundwork},
  volume = {52},
  ISSN = {0010-0277},
  url = {http://dx.doi.org/10.1016/0010-0277(94)90065-5},
  DOI = {10.1016/0010-0277(94)90065-5},
  number = {2},
  journal = {Cognition},
  publisher = {Elsevier BV},
  author = {Goldstone,  Robert L.},
  year = {1994},
  month = aug,
  pages = {125–157}
}

@misc{muttenthaler_2023_improving,
  doi = {10.48550/ARXIV.2306.04507},
  url = {https://arxiv.org/abs/2306.04507},
  author = {Muttenthaler,  Lukas and Linhardt,  Lorenz and Dippel,  Jonas and Vandermeulen,  Robert A. and Hermann,  Katherine and Lampinen,  Andrew K. and Kornblith,  Simon},
  title = {Improving neural network representations using human similarity judgments},
  publisher = {arXiv},
  year = {2023},
  copyright = {Creative Commons Attribution 4.0 International}
}

@article{fu_2023_dreamsim,
  title        = "{DreamSim}: Learning new dimensions of human visual
                  similarity using synthetic data",
  author       = "Fu, Stephanie and Tamir, Netanel and Sundaram, Shobhita and
                  Chai, Lucy and Zhang, Richard and Dekel, Tali and Isola,
                  Phillip",
  year         =  2023,
  primaryClass = "cs.CV",
  eprint       = "2306.09344",
  journal      = "arXiv",
  doi          = "https://doi.org/10.48550/arXiv.2306.09344"
}

@article{dima_simjudg_2022,
    author = {D. C. Dima and T. M Tomita and C. J. Honey and L. Isik},
    title = {Social-affective features drive human representations of observed actions},
    journal = {eLife},
    year = {2022},
    volume = {11},
    issue = {e75027},
    doi = {10.7554/eLife.75027}
}

@article{Hebart2020,
  title = {Revealing the multidimensional mental representations of natural objects underlying human similarity judgements},
  volume = {4},
  ISSN = {2397-3374},
  url = {http://dx.doi.org/10.1038/s41562-020-00951-3},
  DOI = {10.1038/s41562-020-00951-3},
  number = {11},
  journal = {Nature Human Behaviour},
  publisher = {Springer Science and Business Media LLC},
  author = {Hebart,  Martin N. and Zheng,  Charles Y. and Pereira,  Francisco and Baker,  Chris I.},
  year = {2020},
  month = oct,
  pages = {1173–1185}
}

@article{mcmahon_hierarchical_2023,
  title = {Hierarchical organization of social action features along the lateral visual pathway},
  volume = {33},
  ISSN = {0960-9822},
  url = {http://dx.doi.org/10.1016/j.cub.2023.10.015},
  DOI = {10.1016/j.cub.2023.10.015},
  number = {23},
  journal = {Current Biology},
  publisher = {Elsevier BV},
  author = {McMahon,  Emalie and Bonner,  Michael F. and Isik,  Leyla},
  year = {2023},
  month = dec,
  pages = {5035--5047.e8}
}

@article{monfortmoments2019,
  title={Moments in Time Dataset: one million videos for event understanding},
  author={Monfort, Mathew and Andonian, Alex and Zhou, Bolei and Ramakrishnan, Kandan and Bargal, Sarah Adel and Yan, Tom and Brown, Lisa and Fan, Quanfu and Gutfruend, Dan and Vondrick, Carl and others},
  journal={IEEE Transactions on Pattern Analysis and Machine Intelligence},
  year={2019},
  issn={0162-8828},
  pages={1--8},
  numpages={8},
  doi={10.1109/TPAMI.2019.2901464},
}

@inproceedings{song2020mpnet,
  title     = {{MPN}et: Masked and Permuted Pre-training for Language Understanding},
  author    = {Song, Kaitao and Tan, Xu and Qin, Tao and Lu, Jianfeng and Liu, Tie-Yan},
  booktitle = {Advances in Neural Information Processing Systems (NeurIPS)},
  volume    = {33},
  pages     = {16857--16867},
  year      = {2020},
  url       = {https://arxiv.org/abs/2004.09297}
}

@article{mcmahon_seeing_2023,
	title = {Seeing social interactions},
	volume = {27},
	issn = {13646613},
	url = {https://linkinghub.elsevier.com/retrieve/pii/S1364661323002486},
	doi = {10.1016/j.tics.2023.09.001},
	language = {en},
	number = {12},
	urldate = {2024-02-07},
	journal = {Trends in Cognitive Sciences},
	author = {McMahon, Emalie and Isik, Leyla},
	month = dec,
	year = {2023},
	pages = {1165--1179},
}

@article{feichtenhofer_slowfast_2018,
	title = {{SlowFast} {Networks} for {Video} {Recognition}},
	journal = {arXiv: 1812.03982},
	author = {Feichtenhofer, Christoph and Fan, Haoqi and Malik, Jitendra and He, Kaiming},
	year = {2018},
}

@article{kay_kinetics_2017,
	title = {The {Kinetics} {Human} {Action} {Video} {Dataset}},
	journal = {arXiv: 1705.06950},
	author = {Kay, Will and Carreira, Joao and Simonyan, Karen and Zhang, Brian and Hillier, Chloe and Vijayanarasimhan, Sudheendra and Viola, Fabio and Green, Tim and Back, Trevor and Natsev, Paul and Suleyman, Mustafa and Zisserman, Andrew},
	year = {2017},
}

@article{kriegeskorte_representational_2008,
	title = {Representational similarity analysis - connecting the branches of systems neuroscience},
	volume = {2},
	doi = {10.3389/neuro.06.004.2008},
	journal = {Frontiers in Systems Neuroscience},
	author = {Kriegeskorte, Nikolaus and Mur, Marieke and Bandettini, Peter},
	year = {2008},
	pages = {4},
}

@misc{netanyahu_phase_2021,
	title = {{PHASE}: {PHysically}-grounded {Abstract} {Social} {Events} for {Machine} {Social} {Perception}},
	shorttitle = {{PHASE}},
	url = {http://arxiv.org/abs/2103.01933},
	doi = {10.48550/arXiv.2103.01933},
	urldate = {2023-01-26},
	publisher = {arXiv},
	author = {Netanyahu, Aviv and Shu, Tianmin and Katz, Boris and Barbu, Andrei and Tenenbaum, Joshua B.},
	month = mar,
	year = {2021},
	note = {arXiv:2103.01933 [cs, stat]},
}

@article{malik_relational_2023,
	title = {Relational visual representations underlie human social interaction recognition},
	volume = {14},
	copyright = {2023 The Author(s)},
	issn = {2041-1723},
	url = {https://www.nature.com/articles/s41467-023-43156-8},
	doi = {10.1038/s41467-023-43156-8},
	language = {en},
	number = {1},
	urldate = {2023-11-11},
	journal = {Nature Communications},
	author = {Malik, Manasi and Isik, Leyla},
	month = nov,
	year = {2023},
	note = {Number: 1
Publisher: Nature Publishing Group},
	pages = {7317},
}

@misc{bertasius_is_2021,
	title = {Is {Space}-{Time} {Attention} {All} {You} {Need} for {Video} {Understanding}?},
	url = {http://arxiv.org/abs/2102.05095},
	doi = {10.48550/arXiv.2102.05095},
	urldate = {2024-04-25},
	publisher = {arXiv},
	author = {Bertasius, Gedas and Wang, Heng and Torresani, Lorenzo},
	month = jun,
	year = {2021},
	note = {arXiv:2102.05095 [cs]},
}

@misc{radford_learning_2021,
	title = {Learning {Transferable} {Visual} {Models} {From} {Natural} {Language} {Supervision}},
	url = {http://arxiv.org/abs/2103.00020},
	doi = {10.48550/arXiv.2103.00020},
	urldate = {2024-04-25},
	publisher = {arXiv},
	author = {Radford, Alec and Kim, Jong Wook and Hallacy, Chris and Ramesh, Aditya and Goh, Gabriel and Agarwal, Sandhini and Sastry, Girish and Askell, Amanda and Mishkin, Pamela and Clark, Jack and Krueger, Gretchen and Sutskever, Ilya},
	month = feb,
	year = {2021},
	note = {arXiv:2103.00020 [cs]},
}

@article{sundaram_when_2024,
  doi = {10.48550/ARXIV.2410.10817},
  url = {https://arxiv.org/abs/2410.10817},
  journal = {arXiv},
  author = {Sundaram,  Shobhita and Fu,  Stephanie and Muttenthaler,  Lukas and Tamir,  Netanel Y. and Chai,  Lucy and Kornblith,  Simon and Darrell,  Trevor and Isola,  Phillip},
  title = {When Does Perceptual Alignment Benefit Vision Representations?},
  publisher = {arXiv},
  year = {2024},
  copyright = {Creative Commons Attribution Non Commercial Share Alike 4.0 International}
}

@article{Muttenthaler2025,
  title = {Aligning machine and human visual representations across abstraction levels},
  volume = {647},
  ISSN = {1476-4687},
  url = {http://dx.doi.org/10.1038/s41586-025-09631-6},
  DOI = {10.1038/s41586-025-09631-6},
  number = {8089},
  journal = {Nature},
  publisher = {Springer Science and Business Media LLC},
  author = {Muttenthaler,  Lukas and Greff,  Klaus and Born,  Frieda and Spitzer,  Bernhard and Kornblith,  Simon and Mozer,  Michael C. and M\"{u}ller,  Klaus-Robert and Unterthiner,  Thomas and Lampinen,  Andrew K.},
  year = {2025},
  month = Nov,
  pages = {349–355}
}

@article{policzer2025one,
  title={The One Where They Brain-Tune for Social Cognition: Multi-Modal Brain-Tuning on Friends},
  author={Policzer, Nico and Braunstein, Cameron and Toneva, Mariya},
  journal={arXiv preprint arXiv:2511.07988},
  year={2025},
  doi={10.48550/arXiv.2511.07988}
}

@article{schad2025vibe,
  title={VIBE: Video-Input Brain Encoder for fMRI Response Modeling},
  author={Schad, Daniel Carlstr{\"o}m and Dixit, Shrey and Keck, Janis and Studenyak, Viktor and Shpilevoi, Aleksandr and Bicanski, Andrej},
  journal={arXiv preprint arXiv:2507.17958},
  year={2025},
  doi={10.48550/arXiv.2507.17958}
}

@misc{muttenthaler2022vice,
  doi = {10.48550/ARXIV.2205.00756},
  url = {https://arxiv.org/abs/2205.00756},
  author = {Muttenthaler,  Lukas and Zheng,  Charles Y. and McClure,  Patrick and Vandermeulen,  Robert A. and Hebart,  Martin N. and Pereira,  Francisco},
  title = {VICE: Variational Interpretable Concept Embeddings},
  publisher = {arXiv},
  year = {2022},
  copyright = {Creative Commons Attribution Share Alike 4.0 International}
}

\newpage
\appendix

\section{Reproducibility Statement}\label{app:reproducibility}
The data we have collected on the odd-one-out similarity judgments (with the canonical 200/50 train-test split), along with human RSMs and video annotations on all 250 videos will be publicly released. The pertinent metrics for all of the models we have evaluated (both pretrained/baseline and finetuned) are provided in detail in \S\ref{methods}, with additional details on RSA and variance partitioning in the Appendix (\S\ref{app:varpart}). Details on the availability for coding material concerning embedding extraction, similarity computation, and model evaluation (both for $R^2$ scores and odd-one-out accuracy) are provided in the Appendix (\S\ref{app:code}). This section also outlines the way by which others could obtain the Moments in Time dataset we used for our analyses. After gaining access, the mapping we used between raw video files and their representative IDs (0-249) will also be available in the code base. 

Furthermore, we will also release the configuration details for our vision models (V-JEPA 2/2.1, TimeSformer, CLIP, and VideoMAE) with \textsc{LoRA} adapters, including the scripts for hybrid, triplet-only, RSA-only, and triplet-budget-matched versions. See \S\ref{methods} for a more high-level description of training details.
Next, the validation and reporting metrics follow the procedure outlined in \S\ref{subsec:auxiliary_eval}. To facilitate this, we will also release the scripts necessary for full pre-processing and training. You can find more information about the UCF101 linear-probe action recognition and social-affective probing experiments in Appendix \S\ref{app:ucf101}). 
Finally, we wish to support both full retraining and more lightweight reproduction for accessibility. Therefore, we will also make available all training and evaluation code, as well as pretrained adapters and precomputed RSMs to reproduce these analyses. 

\section{Ethics Statement}\label{app:ethics}
All procedures conducted throughout this paper adhere to ethical standards. The behavioral data used was collected under the internal Institutional Review Board (IRB) approval. Informed consent was obtained from all subjects before their participation. The experiment itself was straightforward:  they were instructed to make quick "odd-one-out" choices on 3-second video clips that showed everyday social interactions without any identifying details. We did not collect any personal data beyond demographic information, all of which were optional. We compensated the participants for their time appropriately, and made sure that their responses were reliable without putting too much strain on them. The data we obtained was not used to infer private characteristics on the participants, only for model-human representational alignment. We commit to releasing the dataset (except the actual videos due to licensing, see Appendix \S\ref{app:code}) and the code to encourage transparency and replication.

\section{LLM Usage}
LLMs were used for minor edits, such as grammar, phrasing, and shortening, throughout this paper. LLMs were not used during the data collection or the analysis stages. No major task, such as ideation, full content generation, or substantive interpretation of results, was delegated to LLMs. All conceptual framing, experimental design, and analytical decisions were carried out by the authors.

\section{Compute \& Resource Use}\label{app:compute}
We report a rough estimate of a little over 130 GPU-hours of compute across all 49 models (22 language encoders, 2 image encoders, 2 VLMs, 12 pretrained video backbones, and 11 finetuned video models). Across this compute, the predominant cost is associated with our finetuning process, with a single V-JEPA 2.1 ViT-B + LoRA hybrid training taking $\sim$2.5 GPU-hours alone (12 epochs on a single A100, \texttt{num-epochs=50}, \texttt{early-stopping=5}). When we include all V-JEPA 2.1 and TimeSformer ablations, and the hybrid finetuning process for the baselines of CLIP and VideoMAE, finetuning as a whole consumed about 110 GPU-hours. The remaining compute can be attributed to the checkpoint evaluation across different tasks, including behavioral RSA ($\sim$17 GPU-h), OOO triplet accuracy ($\sim$2 GPU-h), the PHASE OOD probe ($\sim$3 GPU-h), and attention-rollout visualizations ($\sim$0.5 GPU-h). All experiments ran on HPC cluster on a node of 4x NVIDIA A100 (80 GB) GPUs.

\section{Triplet Selection Algorithm}\label{app:triplet_selection_algorithm}

$\mathbf{S}^{(human)}$ is an estimate (aggregate), rather than a fully observed matrix. This is because the triplet sample is sparse relative to all pairs $\binom{250}{2} = 31,125$. We came up with a specialized algorithm to create the triplets to ensure adequate coverage with the least amount of participants possible. So, we designed this procedure so that \emph{every possible pair} $(i, \ j)$ \emph{appears in at least one triplet} $(i, \ j, \ k)$. Since the similarity (probability) matrix is constructed through how many times a pair of videos was rated similar based on how many times they appeared, this guarantees that each “pair” would have at least one rating.

This problem is conceptually equivalent to a \emph{set cover}. Namely, the universe of elements consists of all video pairs, and each triplet corresponds to a subset that covers three of those video pairs. Finding the truly minimal set of triplets that covers all possible pairs is NP-hard. So, we implemented a greedy approximation strategy to iteratively choose the most informative triplet at each step, given all the triplets selected before and remaining.
\begin{itemize}
   \item First, we randomly sample a candidate pool for the triplets at each iteration.
   \item Then, from this pool, we select the triplet that covers the largest number of pairs (max: 3) \emph{not yet included}.
    \item Next, we mark those pairs as ``covered’’ and continue the iteration until every pair has been assigned at least one triplet.
\end{itemize}

We prioritized efficiency with this ``greedy’’ search, since we effectively minimized the number of triplets (and thus the number of participants) we needed to guarantee full pairwise coverage. After coverage was achieved, we adjusted the total number of triplets by adding more so that it was divisible by 220 (11 sets of 20 trials for each participant).

\[
\text{Minimum possible triplets} = \frac{\binom{250}{2}}{3}
= \frac{31{,}125}{3}
= 10{,}375.
\]

\[
\text{Our greedy algorithm produced } 10{,}780 \text{ triplets} \rightarrow
\frac{10{,}780}{220} = 49 \text{ participants required.}
\]

\begin{algorithm}[H]
\caption{Triplet Selection Covering All Pairs (Greedy Set Cover Approximation)}
\label{alg:triplet_selection}
\begin{algorithmic}[1]
\footnotesize{\Require Number of items $N$ (e.g., $N=250$ for 250 video stimuli)
\Ensure Set of triplets $T$ covering all pairs, with $|T|$ divisible by 220
\State $P \gets \{(i,j) \mid 0 \leq i < j < N \}$ \Comment{All pairs}
\State $S \gets \{(i,j,k) \mid 0 \leq i < j < k < N \}$ \Comment{All triplets}
\State $T \gets \emptyset$ \Comment{Selected triplets}

\While{$P \neq \emptyset$}
    \State $C \gets$ random sample of $\min(|S|, 10,000)$ triplets from $S$
    \State $best\_triplet \gets$ triplet in $C$ maximizing coverage w.r.t. $P$
    \State $T \gets T \cup \{best\_triplet\}$
    \State Remove all pairs in $best\_triplet$ from $P$
\EndWhile

\State $r \gets |T| \bmod 220$
\If{$r \neq 0$}
    \State Sample $220 - r$ triplets randomly from $S$ and add to $T$
\EndIf}

\State \Return $T$
\end{algorithmic}
\end{algorithm} 

\section{Supplementary Evaluation and Analysis Procedures}\label{app:supplementary}
\subsection{RSA objective}\label{app:rsa-diff}
During training we use Pearson-correlation RSA on z-scored pairwise distances.
Pearson is smooth, so gradients propagate from the correlation through distances
back to the embeddings. (For evaluation we report Spearman \(\rho^2\) (reported as $R^2$), which is
rank-based and non-differentiable.)

\subsection{On the Hybrid Loss.}\label{app:hybrid_loss}
Our fine-tuning objective combines a triplet loss with an RSA loss, balancing local and global alignment. The triplet component ensures that fine-grained distinctions from the original model are preserved while pulling together pairs judged similar by humans. With the addition of the RSA component, it is complemented by aligning the model's overall pairwise structure with human RSMs. This distills the relational knowledge at a global level, reflecting the findings by \citet{muttenthaler_2023_improving}, who showed that constraints on global geometry that match human similarity yields better, and more interpretable task-effective features when preserving local structure. Where RSA is typically used as a tool for analysis \citep{kriegeskorte_representational_2008}, our contribution takes it a step further and re-purposes it as a differentiable objective. With this, the hybrid loss leverages both local and global supervision to push the representation towards richer semantic space that is reflected by human judgments.

\subsection{Pretrained Model Details}\label{app:model_details}

\paragraph{Video encoders.}
We evaluate nine video encoders spanning CNN and Transformer architectures: X3D-M, X3D-S, and X3D-XS from the X3D family optimized for efficient video classification \citep{feichtenhofer_2020_x3d}; SlowFast-R50, a two-pathway CNN capturing both slow and fast temporal dynamics \citep{feichtenhofer_slowfast_2018}; Slow-R50 and C2D-R50 as single-pathway CNN baselines; I3D-R50; TimeSformer, a video Transformer factorizing spatial and temporal attention trained on Kinetics-400 \citep{kay_kinetics_2017, bertasius_is_2021}; and VideoMAE-Base, a masked autoencoder for self-supervised video pretraining \citep{tong2022videomae}. We feed each 3s clip into these models after resizing frames to the required model resolution and extract embeddings at every layer using the DeepJuice software package 
\citep{Conwell2024_deepjuice}.

\paragraph{Image encoders.}
We benchmark two image models: CLIP ViT-B/32 (vision encoder only) \citep{radford_learning_2021} and DINO-ViT-B/16. For each video, we extract seven equally spaced frames, compute per-frame embeddings, and average across frames before continuing with the standard evaluation 
pipeline.

\paragraph{Vision-language models.}\label{app:vlm_details}
We include Qwen3-VL-2B-Instruct as a modern multimodal reference point, selected for approximate parameter-count parity with V-JEPA~2. We input both the video caption and frames (using the frame-extraction above) and extract embeddings for RSA evaluation.

\paragraph{Dimensionality standardization.}\label{app:jl}
For fairness across architectures with varying hidden dimensionalities, we down-sample embeddings using sparse random projection (SRP) based on the Johnson--Lindenstrauss lemma with $\varepsilon = 0.1$. This automatically sets the projection size according to the number of samples, yielding 4,732 dimensions for the training split ($N=200$) and 3,353 dimensions for the test split ($N=50$), preserving pairwise distances within $\pm 10\%$ with high probability.

\subsection{Variance partitioning analysis}\label{app:varpart}
We model human distances \(d_{\text{human}}(i,j)\) with multiple regression using
model distances as predictors. For models \(X_1, X_2, \ldots\), we fit
\[
\hat d(i,j) \;=\; \beta_0 \;+\; \sum_m \beta_m\, d_{X_m}(i,j)
\]
over all video pairs in the test split, and report \(R^2\).
Unique and shared contributions are obtained by comparing nested models
(e.g., unique \(X_1\) is \(R^2_{X_1,X_2}-R^2_{X_2}\)). We use the best language model as one predictor,
and the pretrained and fine-tuned V-JEPA 2.1 as the other predictors.

\subsection{Split–half reliability}\label{app:splithalf}
We estimate a noise ceiling for the human RSM with a split–half procedure that respects unequal judgments per pair.
In each of 1{,}000 iterations we: (1) restrict to lower-triangle pairs with at least two ratings; (2) reconstruct binary votes (``similar''-``dissimilar'') for each pair using its observed proportion and count, shuffle, and split the votes into two halves; (3) compute the proportion ``similar'' in each half for every pair and take the Spearman correlation across pairs between halves; (4) average these correlations over iterations and apply the Spearman--Brown correction to estimate full-sample reliability. We report this corrected average as the split--half noise ceiling for the human judgments. In figures, we label this as \emph{split--half} \(R^2\), i.e., the squared Spearman--Brown--corrected split--half correlation.

\section{Code and data availability}\label{app:code}

All code used in this paper and our sentence captions are provided for review in \url{https://github.com/garciakathy/bgs-finetuning}. The videos shown to participants for the triplet OOO similarity judgments task are from the Moments in Time (MiT) dataset (\url{http://moments.csail.mit.edu}). The MiT license restricts public release of videos from the dataset, and so we ask to please contact the authors for access.

\section{Social-Affective Behavioral Encoding}
\label{app:beh_encoding}

\begin{figure}[H]
    \centering
    \vspace{1em}
    \includegraphics[width=0.9\textwidth]{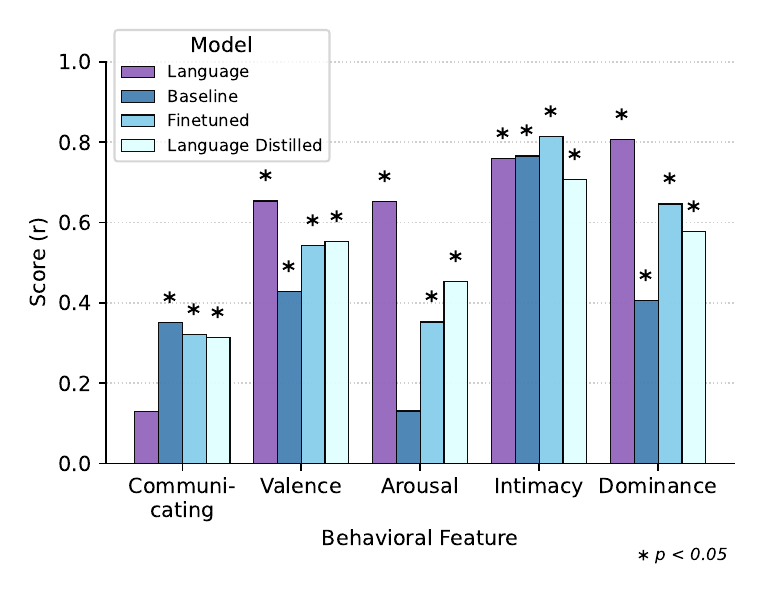}
    \caption{Ridge-regression prediction of social-affective attributes from stimulus embeddings, reported as Pearson correlation ($r$). MPNet provides the language-model comparison; V-JEPA-2.1 variants include baseline, fine-tuned, and language-distilled models. *: $p < 0.05$, permutation test with 5,000 shuffled held-out ratings.}
    \label{behavior_encoding}
\end{figure}

\newpage
\section{Action Recognition Performance}
\label{app:ucf101}

We include here the full results of the UCF101 linear-probe evaluation. All backbone parameters were frozen, and a linear classifier was trained on top of features extracted from the pretrained and fine-tuned TimeSformer and V-JEPA 2.1 models (\texttt{[CLS]} token for TimeSformer; mean-pooled patch tokens for V-JEPA 2.1). Training was repeated across three random seeds, and Top-1 accuracy is reported as mean $\pm$ standard deviation.

\begin{table}[h]
\centering
\caption{Linear probe Top-1 accuracy (\%) on UCF101 split1 with frozen backbones. 
Reported as mean $\pm$ standard deviation over 3 seeds.}
\label{tab:ucf101_probe}
\vspace{0.5em}
\begin{tabular}{lc}
\toprule
Backbone        & Top-1 (\%) \\
\midrule
Pretrained TimeSformer      & $95.75 \pm 0.18$ \\
Fine-tuned TimeSformer      & $95.70 \pm 0.14$ \\
Pretrained V-JEPA-2.1      & $78.17 \pm 0.03$ \\
Fine-tuned V-JEPA-2.1      & $80.31 \pm 0.06$ \\
\bottomrule
\end{tabular}
\end{table}

\section{PHASE Zero Shot Accuracy}
\label{app:phase}

\begin{figure}[H]
    \centering
    \includegraphics[width=0.7\linewidth]{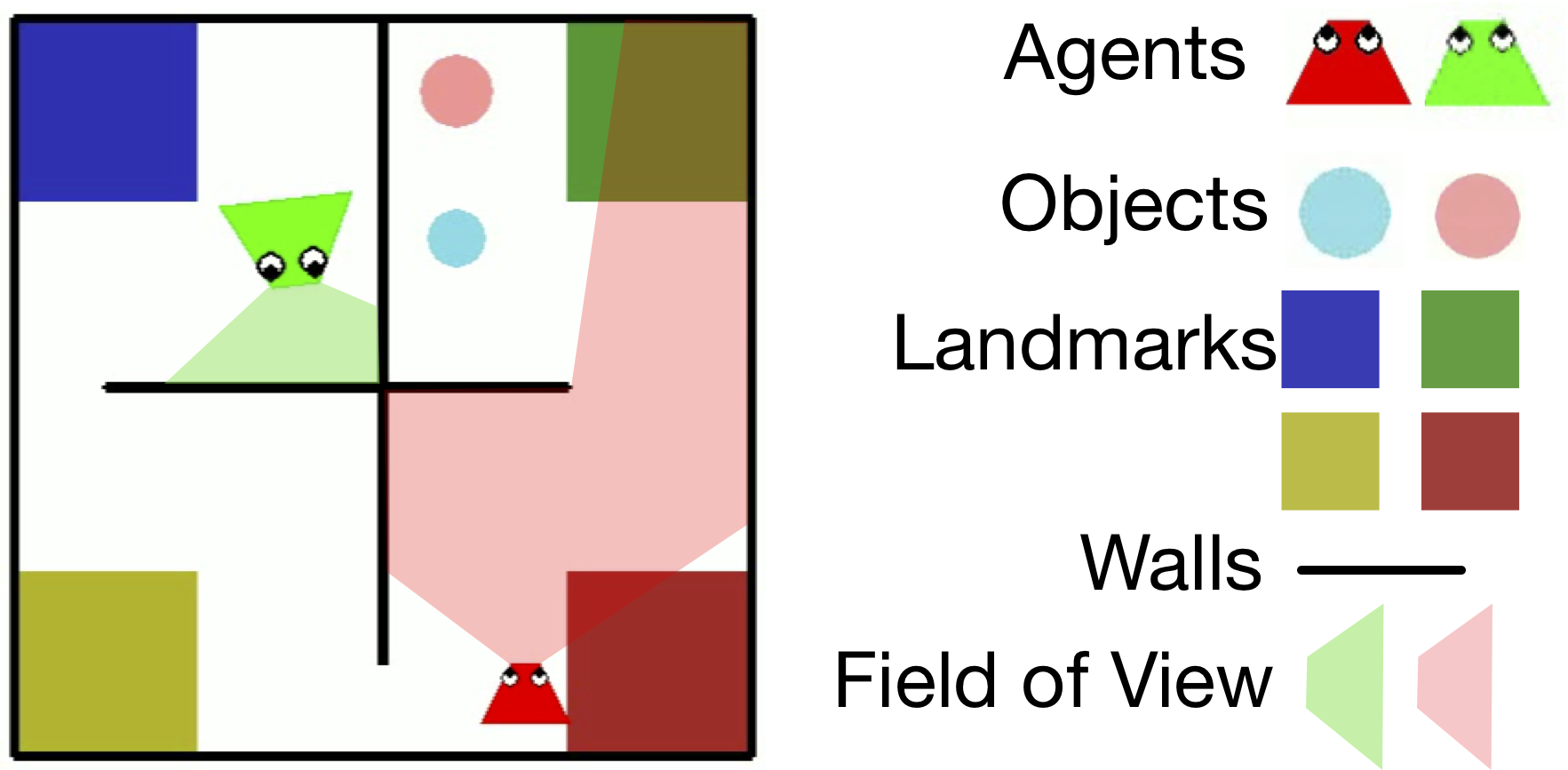}
    \caption{Representative stimuli from the PHASE dataset~\citep{netanyahu_phase_2021}. Example frame shows two agents in a trial (colored triangles) interacting within a 2D environment with walls, obstacles, and goal objects. Trials span a range of social interaction categories including collaborate, help, hinder, follow, chase, and independent, which differ in the agents' goals and the contingency between their movements.}
    \label{fig:phase_stimuli}
\end{figure}

\begin{figure}[H]
    \centering
    \includegraphics[width=0.85\linewidth]{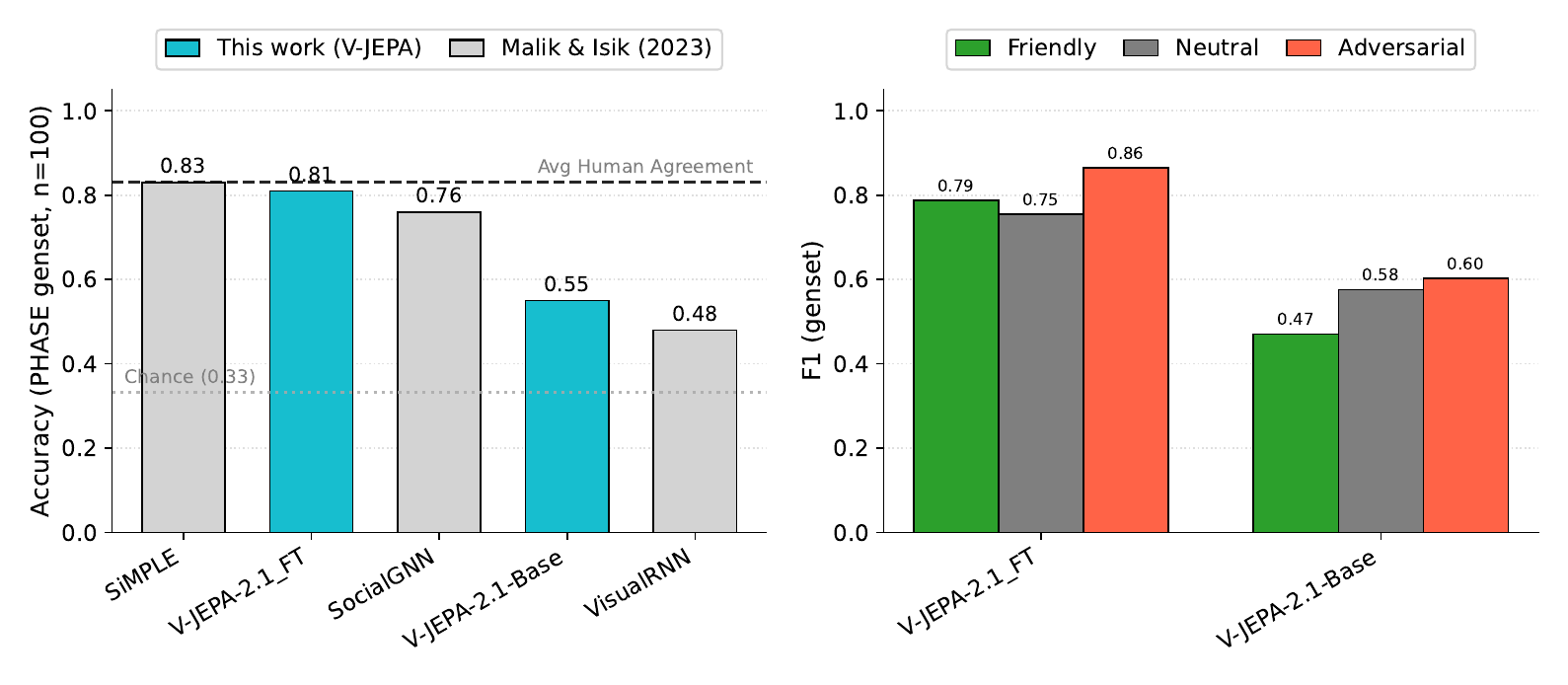}
    \caption{PHASE cross-domain transfer. (Left) 3-way social-interaction classification accuracy on the 100 video held-out generalization set. BGS-aligned V-JEPA 2.1 ViT-B (blue) reaches 0.81 (Wilson 95\% CI: 0.72–0.87), exceeding SocialGNN (0.76) which uses privileged ground truth scene graph input, and matching the hand designed SiMPLE physics grounded inverse planning baseline (0.83) and average human agreement ($\sim$0.83) within sampling error. The pretrained V-JEPA 2.1 baseline (0.55) and the visual recurrent baseline (VisualRNN, 0.48) are substantially lower. Dotted line: chance (33\%). (Right) Per-class F1 for V-JEPA 2.1 ViT-B before and after BGS. Improvement is uniform across friendly, neutral, and adversarial classes, with the strongest gain on adversarial interactions.}
    \label{fig:phase_summary}
\end{figure}

\paragraph{Prior baselines.} The three prior baselines reported in Fig.~\ref{fig:phase_summary} are: \textbf{SocialGNN}~\citep{malik_relational_2023}, a graph neural network that operates on \emph{privileged} symbolic inputs (ground-truth agent positions, velocities, and pairwise distances) extracted from the PHASE simulator rather than on raw pixels; \textbf{VisualRNN}~\citep{malik_relational_2023}, a recurrent visual baseline that processes PHASE clips from raw pixels (no access to the underlying scene graph) and serves as the published pixel-level point of comparison; and \textbf{SiMPLE}~\citep{netanyahu_phase_2021}, a hand-designed, physics-grounded inverse-planning model introduced alongside the PHASE benchmark that infers latent goals and relations via Bayesian inverse simulation rather than learned representations. Reported numbers for SocialGNN and VisualRNN are taken from \citet{malik_relational_2023}; the SiMPLE number and the human agreement estimate are from \citet{netanyahu_phase_2021}.

\section{TimeSformer Finetuned Variance Partitioning}
\label{app:timesformer_var_part}

\begin{figure}[H]
    \centering
    \includegraphics[width=0.75\linewidth]{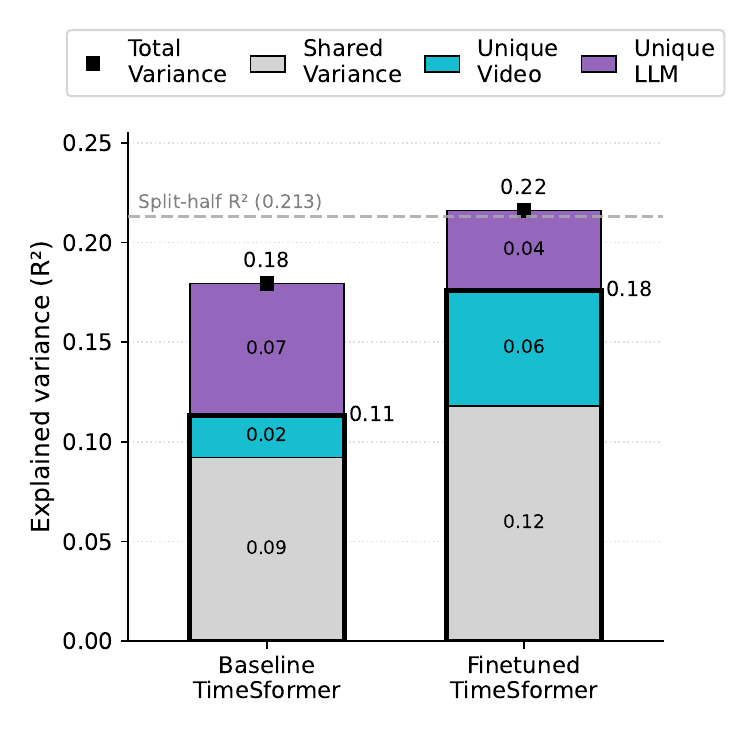}
    \caption{Variance partitioning between TimeSformer and MPNet before and after finetuning. Finetuning shifts explained variance from language-specific to shared and video-specific structure, bringing total variance near the split-half reliability ceiling. Black outlines indicate total variance explained by video-model.}
    \label{fig:model_timesformer_partitioning}
\end{figure}

\newcommand{\NA}{\multicolumn{1}{c}{\textemdash}} 

\newpage
\section{Model Performance and Supervision Budget}\label{secA3}
\begin{table}[H]
  \centering
  \caption{Model performance and supervision constraints budget (— indicates not applicable).}
  \label{tab:perf-budget}
  \small
  \begin{tabular}{lccc}
    \toprule
    \textbf{Model UID} & \textbf{Explained Variance} ($R^2$) & \textbf{OOO Accuracy} & \textbf{Constraints/epoch} \\
    \midrule
    \multicolumn{4}{l}{\textit{Finetuned Models}} \\ 
    \addlinespace[1pt]
    vjepa2-vitl-k710-ft & 0.172117 & 71.67\% & 12978 \\
    vjepa2.1-vitb-dist-ft & 0.164935 & 70.65\% & 12978 \\
    timesformer-ft-hybrid & 0.162023 & 74.46\% & 12978 \\
    vjepa2.1-triplet-only-ft & 0.158260 & 69.29\% & 12978 \\
    timesformer-ft-triplet-match & 0.156857 & 66.58\% & 12978 \\
    clip-vit-b-32-hybrid-finetuned & 0.155953 & 67.84\% & 12978 \\
    timesformer-ft-triplet & 0.145600 & 70.65\% & 12240 \\
    vjepa2.1-rsa-only-ft & 0.131134 & 65.48\% & 13038 \\
    videomae-hybrid-finetuned & 0.123535 & 64.56\% & 12978 \\
    vjepa2.1-mpnet-distilled & 0.123167 & 67.21\% & 12978 \\
    timesformer-ft-rsa & 0.121153 & 63.86\% & 13038 \\
    \addlinespace[4pt]
    \multicolumn{4}{l}{\textit{Video Models}} \\
    \addlinespace[2pt]
    x3d-m & 0.123559 & 68.48\% & \NA \\
    x3d-s & 0.105202 & 64.67\% & \NA \\
    x3d-xs & 0.103721 & 64.95\% & \NA \\
    timesformer-base & 0.102408 & 63.59\% & \NA \\
    i3d-r50 & 0.094969 & 67.66\% & \NA \\
    c2d-r50 & 0.090121 & 65.76\% & \NA \\
    vjepa2-vitl-k710 & 0.086683 & 60.05\% & \NA \\
    videomae-base & 0.086675 & 66.75\% & \NA \\
    slow-r50 & 0.086501 & 67.93\% & \NA \\
    slowfast-r50 & 0.085466 & 64.95\% & \NA \\
    vjepa2.1-vitb-dist & 0.058311 & 63.85\% & \NA \\
    vjepa1-vitl-meanpool-224-16f & 0.052311 & 57.60\% & \NA \\
    \addlinespace[4pt]
    \multicolumn{4}{l}{\textit{Language Models}} \\ 
    \addlinespace[2pt]
    paraphrase-multilingual-mpnet-base-v2 & 0.134374 & 70.38\% & \NA \\
    mxbai-embed-2d-large-v1 & 0.122445 & 66.58\% & \NA \\
    paraphrase-multilingual-MiniLM-L12-v2 & 0.120615 & 67.39\% & \NA \\
    distiluse-base-multilingual-cased-v1 & 0.110899 & 64.95\% & \NA \\
    paraphrase-MiniLM-L6-v2 & 0.102647 & 65.49\% & \NA \\
    all-distilroberta-v1 & 0.101303 & 63.04\% & \NA \\
    stsb-distilroberta-base-v2 & 0.098953 & 64.13\% & \NA \\
    mxbai-embed-large-v1 & 0.090592 & 67.39\% & \NA \\
    all-roberta-large-v1 & 0.088598 & 63.04\% & \NA \\
    all-mpnet-base-v1 & 0.086371 & 66.58\% & \NA \\
    all-mpnet-base-v2 & 0.085562 & 64.67\% & \NA \\
    all-MiniLM-L6-v1 & 0.078124 & 65.22\% & \NA \\
    all-MiniLM-L6-v2 & 0.077037 & 65.49\% & \NA \\
    multi-qa-MiniLM-L6-cos-v1 & 0.068142 & 64.40\% & \NA \\
    all-MiniLM-L12-v2 & 0.065997 & 67.39\% & \NA \\
    LaBSE & 0.052770 & 61.96\% & \NA \\
    clip-ViT-B-32-multilingual-v1 & 0.052506 & 62.77\% & \NA \\
    FacebookAI/roberta-base & 0.025612 & 59.24\% & \NA \\
    FacebookAI/xlm-roberta-base & 0.022418 & 49.46\% & \NA \\
    FacebookAI/roberta-large-mnli & 0.016395 & 47.83\% & \NA \\
    FacebookAI/xlm-roberta-large & 0.010090 & 57.07\% & \NA \\
    \addlinespace[4pt]
    \multicolumn{4}{l}{\textit{Image Models}} \\ 
    \addlinespace[2pt]
    clip-vit-b-32 & 0.126510 & 69.38\% & \NA \\
    dino-dino-vitb16 & 0.075451 & 60.77\% & \NA \\
    \addlinespace[4pt]
    \multicolumn{4}{l}{\textit{Vision-Language Models}} \\ 
    \addlinespace[2pt]
    Qwen3-VL-2B-Instruct & 0.128127 & 68.75\% & \NA \\
    lfm2-5-vl-450m & 0.077825 & 65.76\% & \NA \\
    \bottomrule
  \end{tabular}
\end{table}

\begin{table}[H]
  \centering
  \caption{Subset -- Finetuned and Baseline Models along with best baseline Video, Language, Image and Vision-Language model performance.}
  \label{tab:perf-budget-small}
  \small
  \begin{tabular}{lcc}
    \toprule
    \textbf{Model UID} & \textbf{Explained Variance} ($R^2$) & \textbf{OOO Accuracy} \\
    \midrule
    \multicolumn{3}{l}{\textit{Finetuned}} \\
    \addlinespace[2pt]
    vjepa2-vitl-k710-ft & 0.172117 & 71.67\% \\
    vjepa2.1-vitb-dist-ft & 0.164935 & 70.65\% \\
    timesformer-ft-hybrid & 0.162023 & 74.46\% \\
    vjepa2.1-triplet-only-ft & 0.158260 & 69.29\% \\
    timesformer-ft-triplet-match & 0.156857 & 66.58\% \\
    clip-vit-b-32-hybrid-finetuned & 0.155953 & 67.84\% \\
    timesformer-ft-triplet & 0.145600 & 70.65\% \\
    vjepa2.1-rsa-only-ft & 0.131134 & 65.48\% \\
    videomae-hybrid-finetuned & 0.123535 & 64.56\% \\
    vjepa2.1-mpnet-distilled & 0.123167 & 67.21\% \\
    timesformer-ft-rsa & 0.121153 & 63.86\% \\
    \addlinespace[4pt]
    \multicolumn{3}{l}{\textit{Baseline Models}} \\
    \addlinespace[2pt]
    clip-vit-b-32 & 0.126510 & 69.38\% \\
    timesformer-base & 0.102408 & 63.59\% \\
    vjepa2-vitl-k710 & 0.086683 & 60.05\% \\
    videomae-base & 0.086675 & 66.75\% \\
    vjepa2.1-vitb-dist & 0.058311 & 63.85\% \\
    vjepa1-vitl-meanpool-224-16f & 0.052311 & 57.60\% \\
    \addlinespace[4pt]
    \multicolumn{3}{l}{\textit{Best Video Model}} \\
    \addlinespace[2pt]
    x3d-m & 0.123559 & 68.48\% \\
    \addlinespace[4pt]
    \multicolumn{3}{l}{\textit{Best Language Model}} \\
    \addlinespace[2pt]
    paraphrase-multilingual-mpnet-base-v2 & 0.134374 & 70.38\% \\
    \addlinespace[4pt]
    \multicolumn{3}{l}{\textit{Best Image Model}} \\
    \addlinespace[2pt]
    clip-vit-b-32 & 0.126510 & 69.38\% \\
    \addlinespace[4pt]
    \multicolumn{3}{l}{\textit{Best Vision-Language Model}} \\
    \addlinespace[2pt]
    Qwen3-VL-2B-Instruct & 0.128127 & 68.75\% \\
    \bottomrule
  \end{tabular}
\end{table}

\noindent\textbf{Matching Constraints.}
Despite the same number of optimizer steps across all approaches, the hybrid objective includes an additional RSA term, introducing a modest number of extra supervision signals (\(\approx 738\) pairwise constraints per epoch) beyond the triplet loss (\(12{,}240\) pairwise constraints). To ensure a fair comparison, we trained a \emph{triplet-only (budget-matched)} variant by adding the same number of extra triplet constraints each epoch. This budget-matched triplet model slightly outperforms standard triplet-only training, confirming that more constraints help. Yet, it still under-performs compared to the hybrid model, indicating that the RSA term contributes qualitatively different information by enforcing global structure beyond what can be achieved by simply adding more triplet comparisons.

\newpage
\section{Additional Methods}

\subsection{Sentence captioning of videos}\label{methods_sentence_captioning}

Sentence captions were used from a publicly available dataset \cite[see][for full details]{garcia2025modeling}. A group of 150 participants was recruited on Prolific to provide sentence captions.  Eligibility criteria for this online study required participants to be native English speakers, 18 years or older (M=39.72, SD=13.24), with normal or corrected-to-normal vision. The cohort consisted of 63 females and 87 males. Self-reported race and ethnicity were as follows: 114 white, 14 black, 10 Asian, 9 mixed race, 2 other, and 3 who declined to report.

\begin{figure}[ht]
    \centering
    \vspace{1em}
    \includegraphics[width=0.9\textwidth]{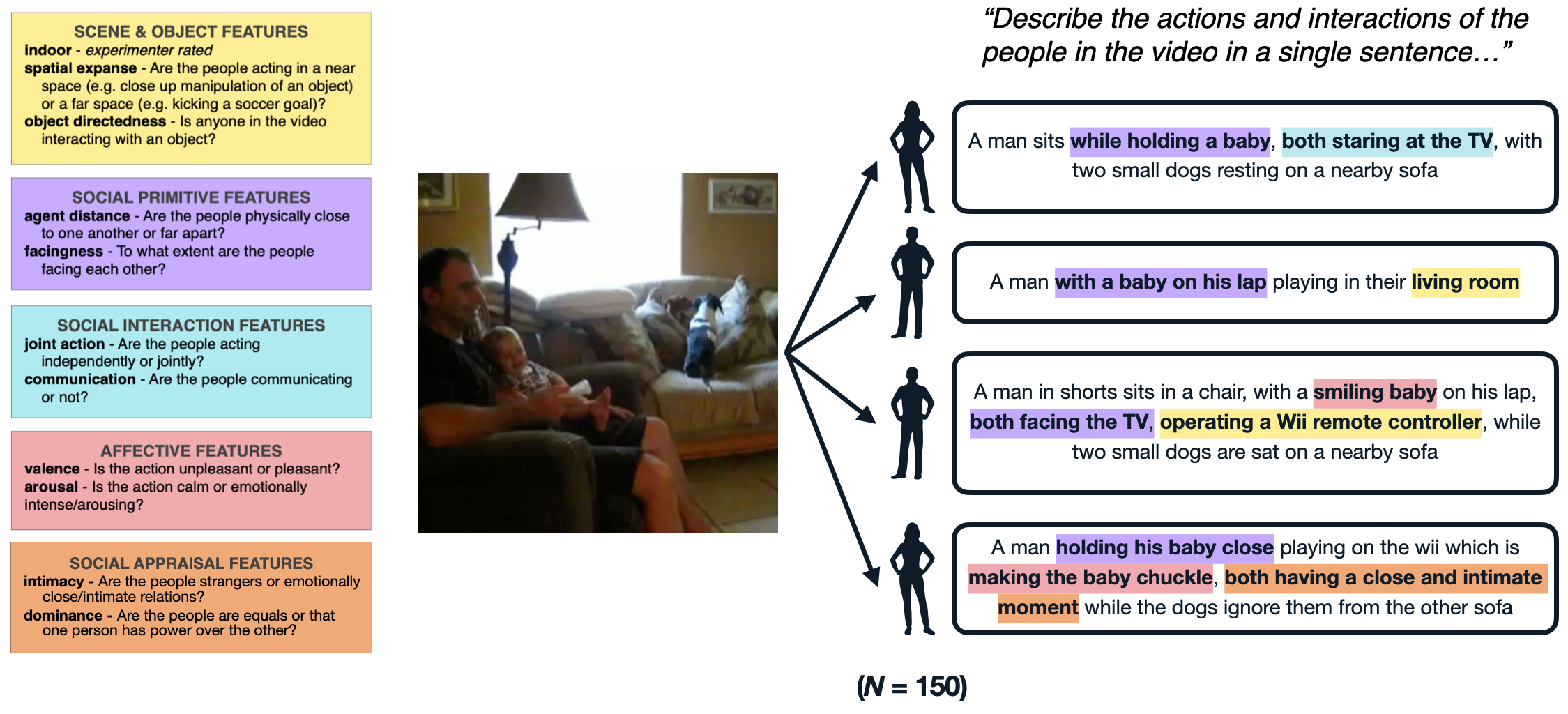}
    \caption{Example video and participant descriptions from the publicly available dataset from \cite{garcia2025modeling}. Participants viewed a short video clip (center) and were asked to describe the actions and interactions of the people in the scene using a single sentence. Example responses (right) show how different aspects of the same video can emphasize distinct feature categories: scene and object features (yellow), social primitive features (purple), social interaction features (blue), affective features (red), and social appraisal features (orange) (Modified from \cite{mcmahon_hierarchical_2023}). Each highlighted phrase corresponds to the feature dimension it represents.}
    \label{fig:sentence_captions}
\end{figure}

The task required each participant to write a single-sentence caption for 12 videos presented in a random order (10 standard and 2 catch trials). Participants typed their responses into a text box that initially showed a placeholder prompt: ``Description of the actions and interactions of the people in the video in a single sentence...'' (Appendix Fig. \ref{fig:sentence_captions}).

\newpage
\subsection{Participant Demographics for Odd-One-Out Similarity Dataset.}
\begin{figure}[H]
    \centering
    \vspace{1em}
    \includegraphics[width=0.95\textwidth]{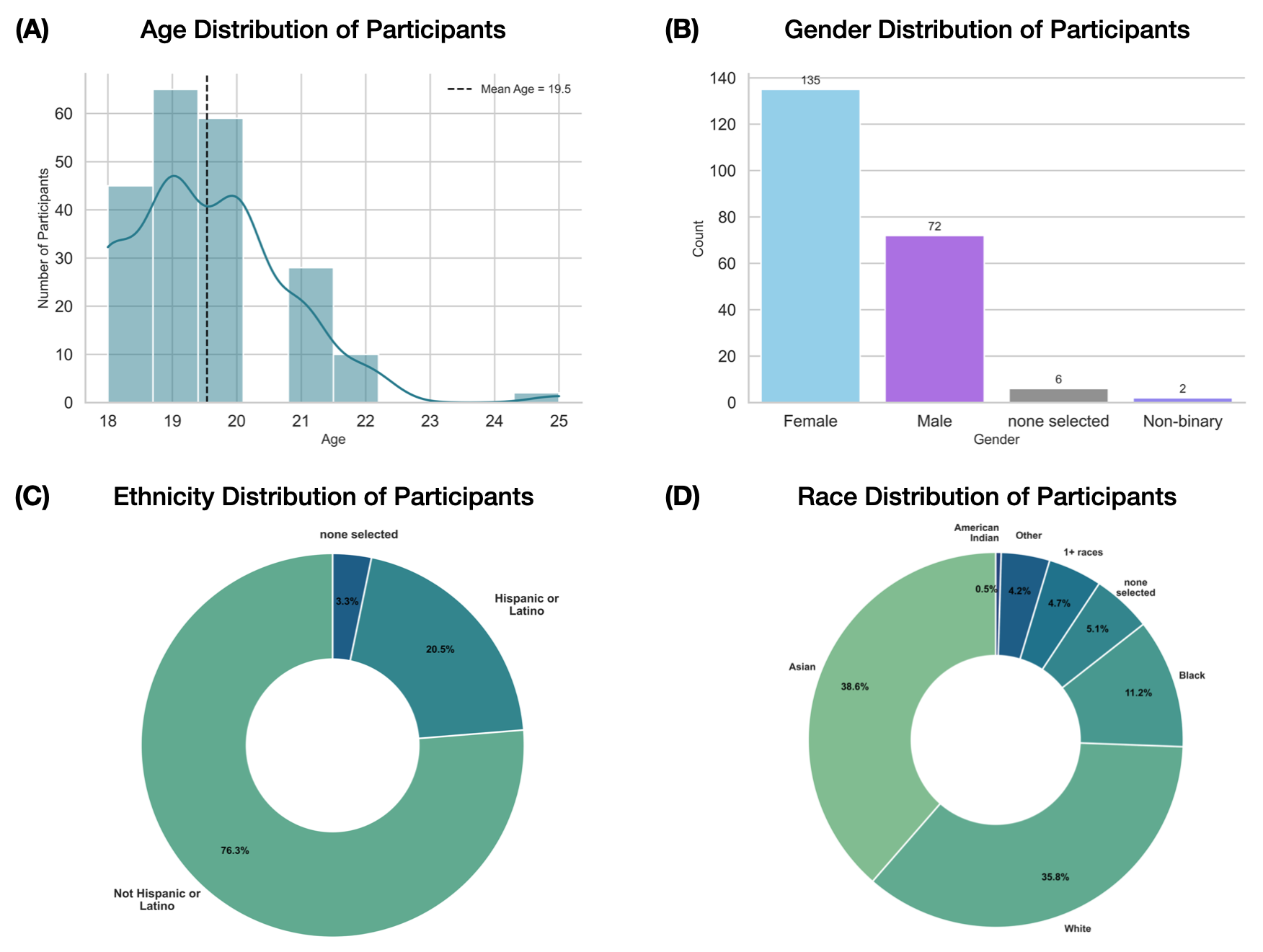}
    \caption{Participant Demographics. (A) Age distribution (mean = 19.5 years). (B) Gender distribution. (C) Ethnicity distribution. (D) Race distribution. All participants were recruited online via our University's psychological research platform, and participated in the study on Meadows Research (\url{https://meadows-research.com}). All participants gave informed consent in accordance with our internal Institutional Review Board, who provided explicit approval of all protocols and procedures discussed. All participants were 18 or older, with normal or corrected-to-normal vision, and at least academically proficient English speakers. }
    \label{fig:demographics}
\end{figure}

\newpage
\subsection{Video Action Category Distribution}

\begin{table}[H]
\centering
\caption{Action category frequencies across all 250 videos, showing the full distribution of annotated behaviors from the most common everyday actions to the least frequent, diversity-focused categories.}
\begin{tabular}{l r @{\hspace{3em}} l r}
\toprule
\textbf{category} & \textbf{count} & \textbf{category} & \textbf{count} \\
\midrule
crying & 21 & building & 2 \\
laughing & 18 & boating & 2 \\
drumming & 18 & baking & 2 \\
brushing & 18 & closing & 2 \\
fishing & 11 & wrestling & 2 \\
dancing & 10 & bowing & 2 \\
giggling & 9 & gardening & 2 \\
cooking & 9 & shopping & 1 \\
discussing & 8 & digging & 1 \\
clapping & 7 & pushing & 1 \\
driving & 6 & unloading & 1 \\
smoking & 6 & exercising & 1 \\
working & 6 & drilling & 1 \\
eating & 6 & applauding & 1 \\
singing & 5 & spitting & 1 \\
planting & 5 & catching & 1 \\
reading & 5 & camping & 1 \\
bathing & 5 & barbecuing & 1 \\
playing & 5 & hugging & 1 \\
kicking & 5 & riding & 1 \\
mowing & 4 & chewing & 1 \\
hiking & 4 & cleaning & 1 \\
throwing & 4 & speaking & 1 \\
skating & 3 & playing+videogames & 1 \\
drinking & 3 & drawing & 1 \\
walking & 3 & skiing & 1 \\
dipping & 3 & jogging & 1 \\
hunting & 3 & studying & 1 \\
knitting & 3 & bowling & 1 \\
 &  & unpacking & 1 \\
\bottomrule
\end{tabular}
\label{tab:action_distribution}
\end{table}

\newpage
\subsection{Proportions of Nouns and Verbs Across Video Captions}
\begin{table}[H]
\centering
\caption{Most frequent nouns and verbs across all captions, showing the proportion of word occurrence across captions.}
\vspace{0.25em}
\begin{tabular}{lrlr}
\toprule
Noun & Noun Proportion & Verb & Verb Proportion \\
\midrule
man & 0.255 & play & 0.173 \\
baby & 0.203 & sit & 0.083 \\
woman & 0.151 & hold & 0.067 \\
child & 0.133 & look & 0.065 \\
girl & 0.083 & talk & 0.052 \\
boy & 0.081 & cry & 0.051 \\
people & 0.058 & watch & 0.046 \\
drum & 0.052 & stand & 0.046 \\
adult & 0.044 & brush & 0.043 \\
hand & 0.041 & laugh & 0.038 \\
tooth & 0.039 & sing & 0.025 \\
toddler & 0.037 & dance & 0.024 \\
water & 0.032 & appear & 0.024 \\
mother & 0.031 & smile & 0.024 \\
kid & 0.026 & try & 0.022 \\
toy & 0.025 & help & 0.021 \\
fishing & 0.025 & feed & 0.021 \\
person & 0.024 & make & 0.020 \\
car & 0.024 & use & 0.019 \\
lady & 0.022 & lie & 0.019 \\
kitchen & 0.022 & have & 0.019 \\
bed & 0.021 & walk & 0.018 \\
book & 0.019 & fish & 0.018 \\
hair & 0.018 & show & 0.017 \\
father & 0.018 & do & 0.017 \\
dance & 0.017 & seem & 0.017 \\
guy & 0.017 & put & 0.017 \\
food & 0.017 & eat & 0.017 \\
friend & 0.017 & lay & 0.016 \\
front & 0.017 & throw & 0.014 \\
floor & 0.016 & move & 0.013 \\
son & 0.015 & take & 0.013 \\
bath & 0.015 & read & 0.013 \\
arm & 0.015 & face & 0.013 \\
game & 0.015 & wear & 0.012 \\
male & 0.014 & speak & 0.012 \\
brother & 0.014 & explain & 0.012 \\
mouth & 0.014 & get & 0.011 \\
guitar & 0.014 & catch & 0.010 \\
ball & 0.014 & work & 0.009 \\
side & 0.013 & learn & 0.009 \\
dog & 0.013 & prepare & 0.009 \\
playing & 0.012 & smoke & 0.009 \\
room & 0.012 & cook & 0.009 \\
dad & 0.012 & interact & 0.009 \\
ice & 0.011 & discuss & 0.009 \\
microphone & 0.011 & drink & 0.009 \\
boat & 0.011 & clean & 0.009 \\
time & 0.011 & drive & 0.009 \\
face & 0.011 & practice & 0.008 \\
\bottomrule
\end{tabular}
\label{tab:noun_verb_props}
\end{table}

\clearpage
\section{Supplemental Qualitative Interpretability}\label{app:supplemental_comparison}

\subsection{Model-Human Agreement for the Same Set of Triplets}

\begin{figure}[H]
    \centering
    \includegraphics[width=0.85\linewidth]{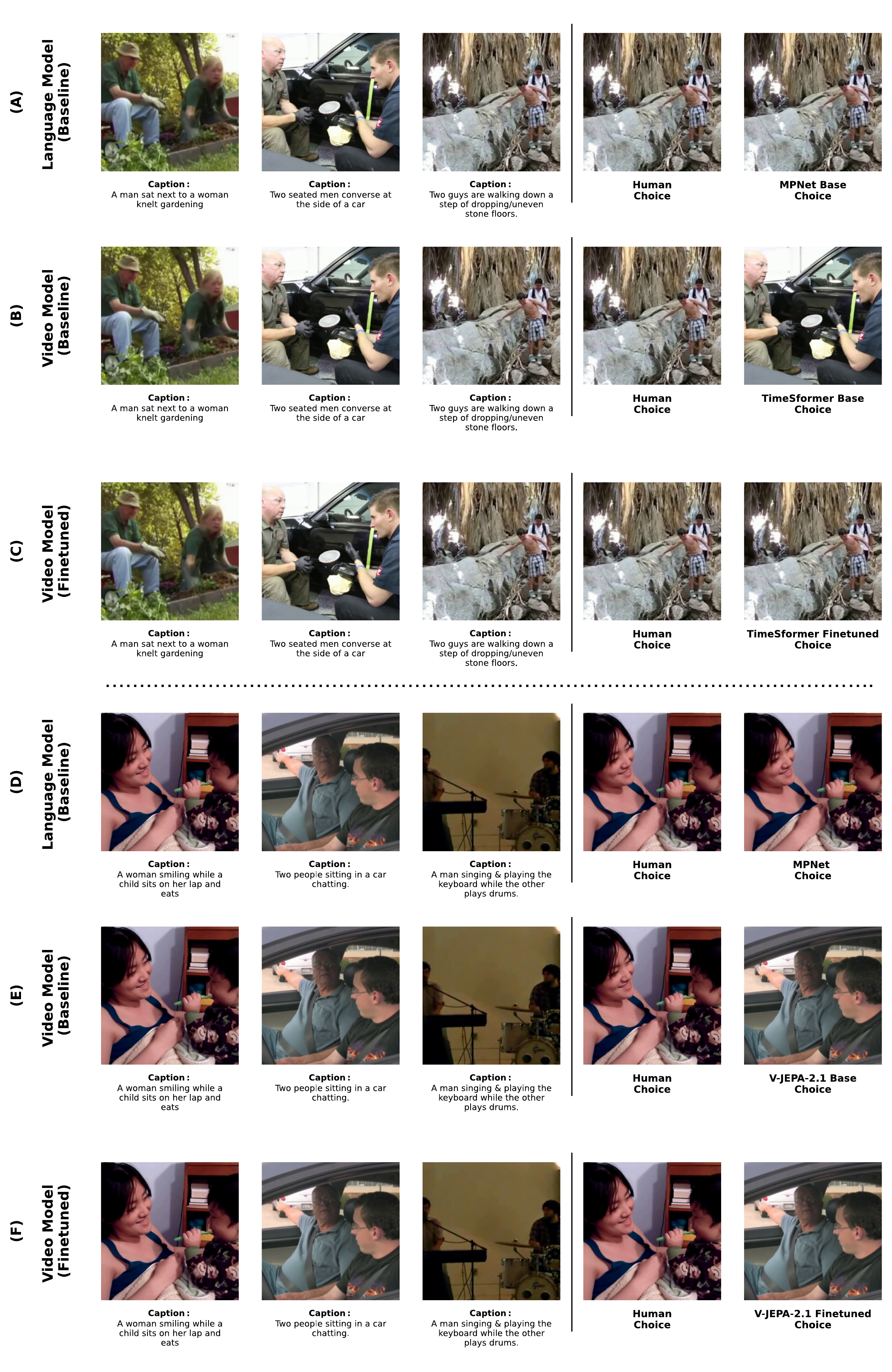}
    \caption{Human–model agreement and disagreement on triplet odd-one-out judgments across three modalities (A, B, C) baseline MPNet, baseline V-JEPA-2.1 and finetuned V-JEPA-2.1 respectively where V-JEPA-2.1 agrees with human judgment after finetuning. (D, E, F) baseline  MPNet, baseline V-JEPA-2.1 and finetuned V-JEPA-2.1 respectively where V-JEPA-2.1 disagrees with human judgment after finetuning.}
    \label{fig:model_human_comparison-single}
\end{figure}

\newpage
\subsection{Qualitative Examples of Human–Model Agreement}
\begin{figure}[ht]
    \centering
    \includegraphics[width=0.85\linewidth]{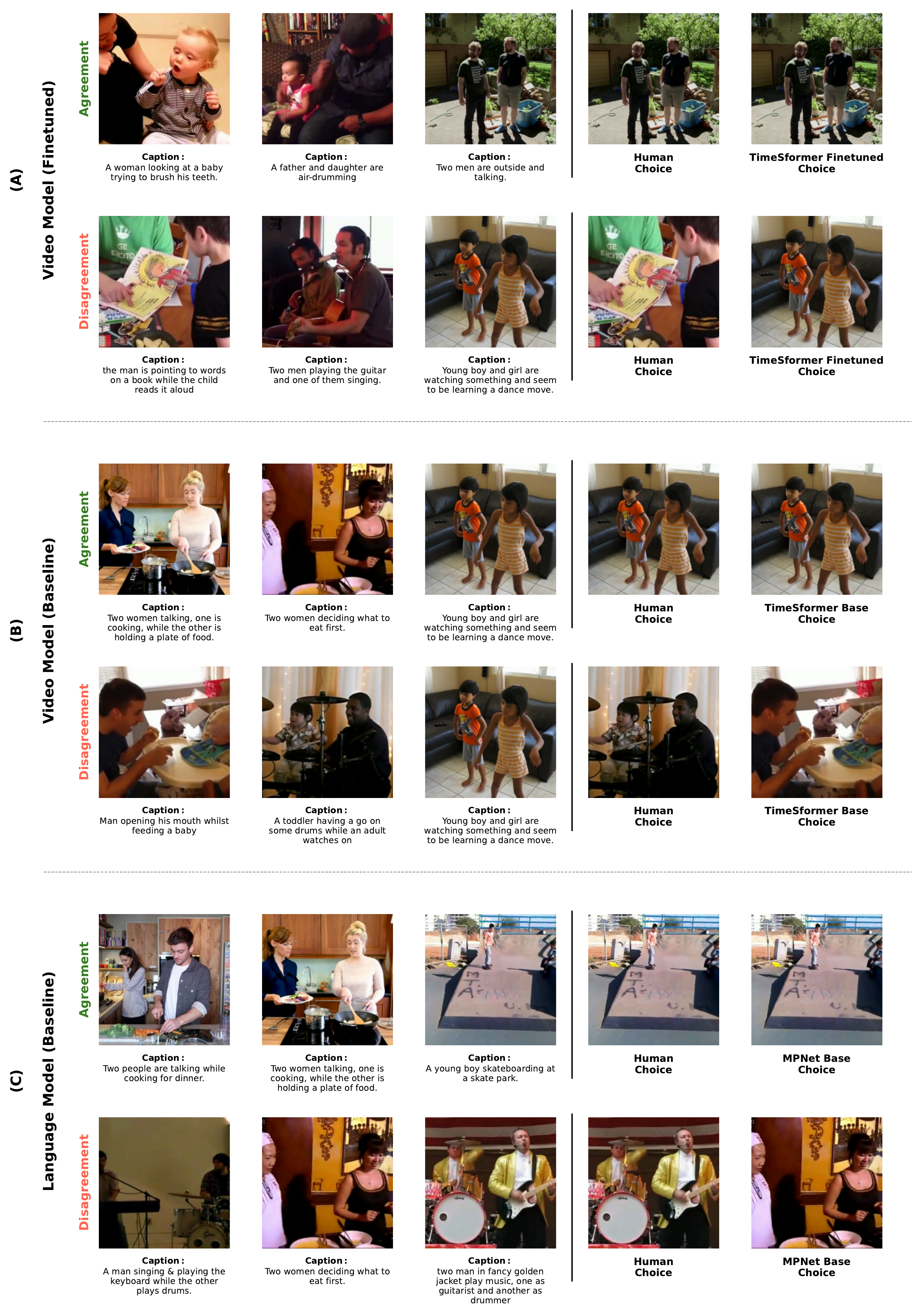}
    \caption{Human–model agreement and disagreement on triplet odd-one-out judgments across three modalities \textit{(A) finetuned TimeSformer}, \textit{(B) baseline TimeSformer}, and \textit{(C) baseline MPNet}, where each row shows the three candidate videos, the human-selected odd one out, the model-selected odd one out, and the human-written captions, with horizontal separators marking modality groups and row labels indicating agreement or disagreement.}
    \label{fig:model_human_comparison}
\end{figure}

\newpage
\subsection{Attention Rollout Comparison - V-JEPA-2.1}

\begin{figure}[H]
    \centering
    \includegraphics[width=0.75\linewidth]{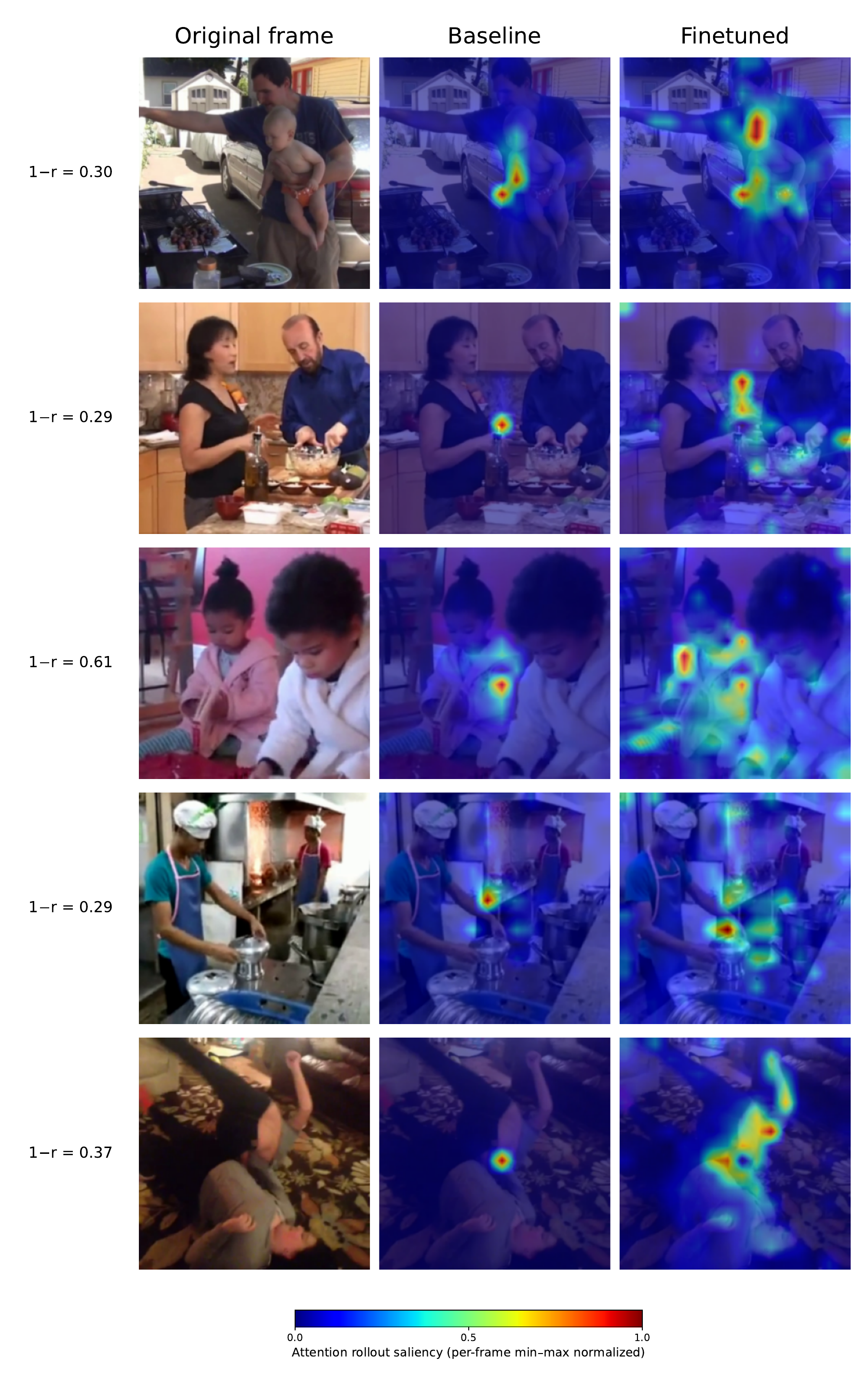}
    \caption{Spatial attention rollout for V-JEPA~2.1 ViT-B before and 
after BGS fine-tuning. We select the five stimuli with the 
largest attention divergence between models, quantified as 
$1 - \bar{r}$: one minus the mean Pearson correlation between baseline 
and fine-tuned per-frame saliency maps, averaged over the model's 8 
temporal slices. \textbf{Left:} original frame. \textbf{Center:} 
pretrained model, which concentrates attention on a single focal region 
(typically one agent's torso or limbs). \textbf{Right:} fine-tuned 
model, which distributes attention across socially informative 
regions---faces, gaze direction, interacting hands, and both 
agents---without any spatial supervision during training. $1 - \bar{r}$ 
values for the five videos shown are $0.30$, $0.29$, $0.61$, $0.29$, 
and $0.37$, compared to the dataset-wide mean of 
$0.18 \pm 0.03$. Even among these maximally divergent cases, the shift 
is consistently toward socially informative regions rather than 
arbitrary redistribution; the narrow dataset-wide variance confirms 
systematic reweighting of spatial priority across the full 250-video 
stimulus set.}
    \label{fig:attention-comparison}
\end{figure}

\subsection{Attention Rollout Comparison - TimeSformer}

\begin{figure}[H]
    \centering
    \includegraphics[width=0.64\linewidth]{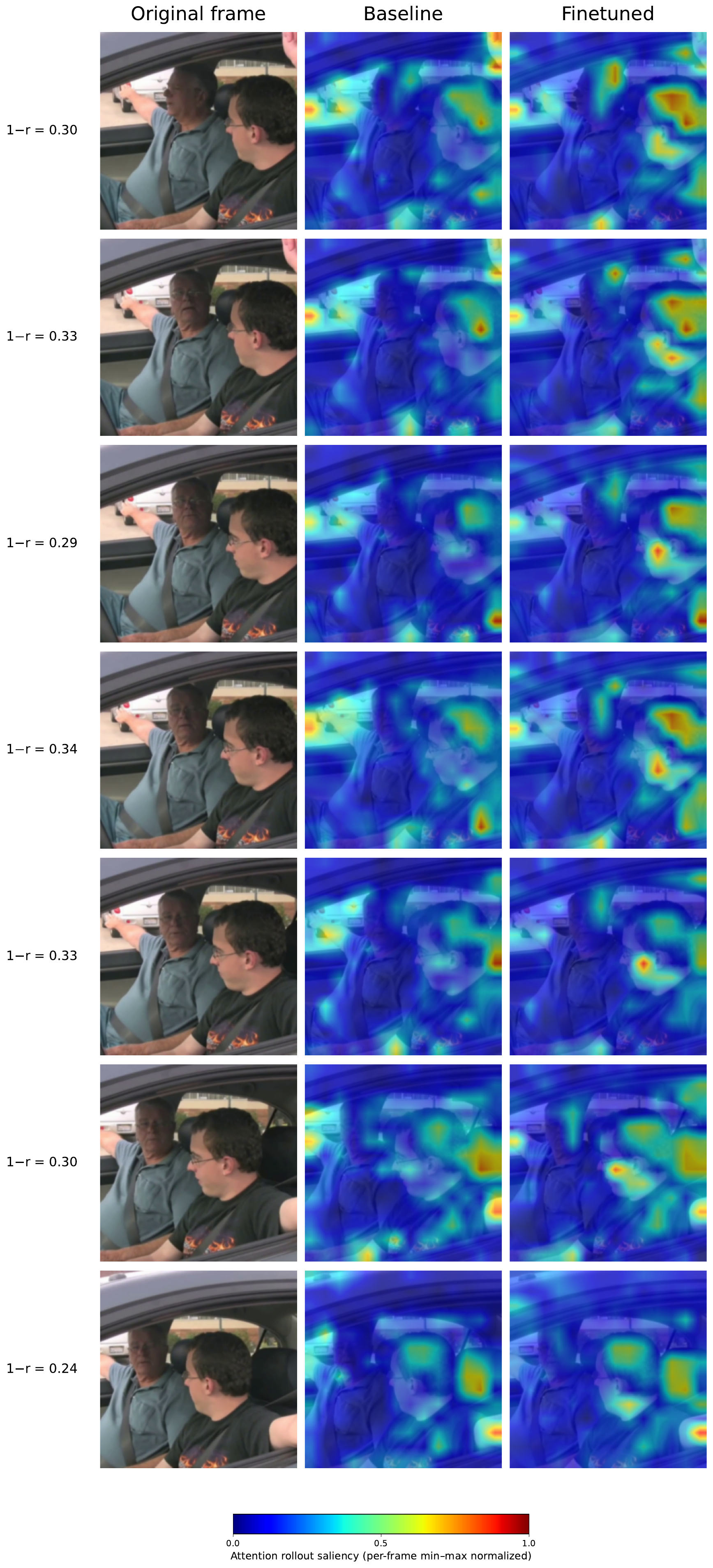}
    \caption{Spatial attention rollout for TimeSformer before and after BGS fine-tuning. We select a representative stimuli and show attention over the model's 8 temporal slices. \textbf{Left:} original frame. \textbf{Center:} pretrained model. \textbf{Right:} fine-tuned model, which shifts attention across socially informative regions---heads, faces, gaze direction, and both agents---without any spatial supervision during training.}
    \label{fig:attention-comparison-timesformer}
\end{figure}

\end{document}